# An Effective Strategy to Enable Argyrodite Sulfides as Superb Solid-State Electrolytes: Safeguarding Remarkable Ionic Conductivity and Interfacial Stability with Electrodes

*Hongjie Xu[1,2,3], Yuran Yu[1,2,3], Junhua Hu[1,2,3]*, Zhuo Wang[1,2,3]*, and Guosheng Shao[1,2,3]**

1 School of Materials Science and Engineering, Zhengzhou University, Zhengzhou 450001, Henan, China
2 State Centre for International Cooperation on Designer Low-carbon &Environmental Materials (CDLCEM), Zhengzhou University, Zhengzhou 450001, Henan, China
3 Zhengzhou Materials Genome Institute (ZMGI), Zhengzhou 450100, Henan, China

E-mail: hujh@zzu.edu.cn; wangzh@zzu.edu.cn; gsshao@zzu.edu.cn

Abstract: The argyrodite sulfides are getting more and more attractive as highly promising solid-state electrolytes (SSEs) for high-performance all-solid-state batteries (ASSBs), owing to their high ionic conductivity, adequate plasticity, and decent mechanical strength. However, their poor incompatibility with Li metal anode and high voltage cathodes and as well as serious sensitivity to air significantly hinder their practical applications. Herein, we have devised an effective strategy to overcome these challenging shortcomings through modification of chalcogen chemistry under the guidance of theoretical modeling. The resultant $Li_{6.25}PS_4O_{1.25}Cl_{0.75}$ delivered excellent electrochemical compatibility with both pure Li anode and high-voltage $LiCoO_2$ cathode, without detrimental impact upon the superb ionic conductivity of the pristine sulfide. Furthermore, the current SSE also exhibited highly improved stability to oxygen and moisture in air, with further advantage being more insulating to electrons. The remarkably enhanced compatibility with electrodes is attributed to in situ formation of solid anode electrolyte interphase (AEI) and cathode electrolyte interphase (CEI) layers. The formation of *in situ* AEI enabled ultra-stable Li plating/stripping at a record high current density up to 1 mAh cm$^{-2}$ in Li|$Li_{6.25}PS_4O_{1.25}Cl_{0.75}$|Li symmetric cells over 1800 hours. The *in situ* CEI facilitated protection of the SSE from decomposition at elevated voltage. Consequently, the synergistic effect of AEI and CEI helped the $LiCoO_2|Li_{6.25}PS_4O_{1.25}Cl_{0.75}|Li$ battery cell to achieve markedly better cycling stability than that using the pristine $Li_6PS_5Cl$ as SSE, at a high areal loading of the active cathode material (4 mg cm$^{-2}$). This work adds a desirable SSE in the argyrodite sulfide family, so that high-performance solid battery cells could even be fabricated in ambient air.

## 1. Introduction

Solid-state electrolytes (SSEs) have attracted great attention owing to the pressing challenge in addressing the safety issues in conventional lithium-ion batteries (LIBs) from flammable organic liquid electrolytes.[1-2] As a landmark breakthrough, ultra-fast transportation of $Li^+$ ions has been achieved in different sulfide compounds, especially in the $Li_{10}GeP_2S_{12}$ (LGPS)[3-4] and the argyrodite $Li_6PS_5Cl$[5-7] systems. Both of them exhibit outstanding $Li^+$ ion conductivities that is comparable to or even much better than those of organic liquid electrolytes.[1, 4, 8]

Sulfide SSEs have many promising merits covering excellent $Li^+$ ion conductivity, less serious grain/particle boundary issues, low processing temperature (around 500 °C), and easiness for fabricating high-performance battery cells owing to their good deformability to enable high-quality bonding between particles.[4, 9-11] However, this class of SSEs still have some serious shortcomings,[12-14] noticeably: (a) sensitive to air and moisture; (b) prone to conductive to electrons due to narrow bandgaps; (c) inadequate stability with lithium metal anode; and (d) incompatible with high-voltage cathode materials.

For sulfide-based SSEs, the chemical instabilities in humid air will inevitably cause deterioration of both structure and composition, leading to significant loss in ionic conductivity and release of noxious $H_2S$ gas.[1] Consequently, their processing and handling have to be strictly isolated from air and humidity, leading to complication in manufacturing with higher cost. Moreover, the reported band gaps of sulfide-based SSEs are only 2 to 3 eV,[15] which make them more likely get less insulating to electrons when deviation from stoichiometry triggers self- or impurity-doping effect, undesirable to battery electrolyte.

Lithium metal is considered as a "Holy Grail" anode material for a rechargeable battery as it has the highest specific capacity of 3860 mA h $g^{-1}$ and the lowest redox potential of −3.04 V ($Li^+$/Li) (*vs.* standard hydrogen electrode, SHE).[16-17] This makes it the most attractive anode for solid batteries with high energy densities. Unfortunately, there exist serious electrochemical incompatible issues between sulfide SSEs and Li.[18] According to theoretical analysis, either the LGPS family or argyrodite-type sulfides exhibit a reduction potential at around 1.7 V *vs.* Li/$Li^+$,[18] and detrimental interphases tend to form at their interfaces with the Li anode. Experimental work also revealed that sulfide-SSEs are prone to decomposition into $Li_xP$ and $Li_2S$ when they are in direct contact with Li metal.[19] Formation of such anode-electrolyte interphases (AEI) was linked to the growth of lithium dendrites that caused short circuiting in ASSBs.[20-22] Rather complex technologies were thus employed to modify the surface of Li, in order to avoid the formation of such undesirable AEI layer at the interface of Li|sulfide-SSEs.[20-24] Alternatively, it can help reduce the decomposition of sulfide-SSEs to employ materials with higher electrochemical potential as anodes e.g. Li–Sn[25-26] or Li–In alloy (0.6 V *vs.* Li/$Li^+$)[19] or $Li_4Ti_5O_{12}$ (1.5 V *vs.* Li/$Li^+$),[27-28] albeit being at the expense of the energy density of ASSBs.

Use of high-voltage cathode material is also key to delivering high energy density ASSBs.[29-30] The current intercalation type cathode materials, such as the well exploited $LiCoO_2$,[31] $LiNi_aCo_bMn_{(1-a-b)}O_2$,[32] $LiNi_{0.8}Co_{0.15}Al_{0.05}O_2$[33] and olivine $LiFePO_4$,[34] have fairly high voltage plateaus above 3.3 – 4.2 V, which are drastically higher than the maximum oxidation potential of sulfide SSEs around 2.1 V.[35] Such big gaps between the reduction potential of cathode and the oxidation potential of SSEs lead to fundamental electrochemical incompatibility at the interfaces between sulfide SSEs and cathode materials,[35-36] and this in turn results in the formation of a Li depletion layer as a space charge region at each cathode-SSE interface [37-38] and thus finally leads to rapid capacity loss of ASSBs due to reactions at the SSE|cathode interfaces.[19, 39-40] Formation of cathode-electrolyte interphases (CEI) with usually poor ionic conductivity significantly blocks the transportation of $Li^+$ during charging-discharging.[41-42] Therefore, functional buffer layers, which can tolerate the high-voltage to prevent detrimental reactions at the cathode-electrolyte interface, have been considered to coat cathode particles to address such interfacial incompatibilities.[35, 43-44] However, most high-voltage buffer layers have rather limited ionic conductivities,[45] so that great efforts have been made to develop rather costly coating technologies to limit coating thickness without compromising full coverage.[46-49] It is envisaged that a more appealing technological route is towards significant improvement of the voltage tolerance of sulfide-SSEs without detrimental effect on their ionic conductivities.

The argyrodite electrolytes $Li_6PS_5X$ (X = Cl, Br, and I) are a highly promising family of sulfide electrolytes, which are based on cheap raw materials with fairly high ionic conductivity and better electrochemical stability than LGPS.[5, 7, 9, 44, 50] The $Li^+$ ion conductivity of $Li_6PS_5Cl$ can be further improved through suitable optimization of chemical composition, such as in the slightly off-stoichiometric composition $Li_{6.25}PS_{5.25}Cl_{0.75}$.[9, 15] A Cl-rich composition $Li_{5.5}PS_{4.5}Cl_{1.5}$ was reported to exhibit a conductivity of 12 mS $cm^{-1}$ at room temperature.[51] However, the compatibility issues with electrodes are not yet fully resolved due to their intrinsically narrow electrochemical window of argyrodite electrolyte. In our recent theoretical modeling, we predicted that the electrochemical window of $Li_6PS_5Cl$ can be extended through partially replacing S with O, which enhances electrochemical stability with both oxide cathodes and Li anode.[35] It was found that the ratio of O and S is the decisive factor to impact on the $Li^+$ conductivity and materials stability. Here in this work, we focus on experimental exploitation to synthesize a new class of electrolytes, via O substitution of S, leading to the discovery of $Li_{6.25}PS_4O_{1.25}Cl_{0.75}$ as a highly attractive SSE that enables greatly improved materials stability and electrochemical compatibility with electrodes but with little detrimental impact upon the high ionic conductivity of the pristine sulfide SSEs. This new SSE is not only compatible with both the high voltage $LiCoO_2$ cathode and Li metal anode, but is also remarkably insensitive to humid air (humidity 53%). The great electrochemical compatibility with high voltage cathode and Li anode is owing to *in situ* formation of solid anode electrolyte interphase (AEI) and cathode electrolyte interphase (CEI) layers, so that the AEI enabled ultra-stable Li plating/stripping at a high current density up to 1 mAh $cm^{-2}$ in Li|$Li_{6.25}PS_4O_{1.25}Cl_{0.75}$|Li symmetric cells over 1800 hours. The combined advantages from *in situ* electrolyte-electrode interphases, AEI and CEI, facilitated the $LiCoO_2$|$Li_{6.25}PS_4O_{1.25}Cl_{0.75}$|Li battery cell to achieve significantly improved cycling stability with respect to that using the pristine $Li_6PS_5Cl$ as SSE, at a high areal loading of the cathode material (4 mg $cm^{-2}$). This work adds another superb SSE in the argyrodite family, thus offering a solid materials basis towards engineering high-performance solid battery cells processible in ambient air.

## 2. Results and Discussion

### 2.1 Energy Convex Hull of $Li_6PS_{5(1-x)}O_{5x}Cl$ and Synthesis

The thermodynamic effect of O substitution of S is assessed by first-principles ATAT simulation, taking room temperature stable compounds $Li_6PS_5Cl$[5, 7] and $Li_6PO_5Cl$[52] as two terminal reference phases. **Figure 1**a shows the energy convex hull of $Li_6PS_{5(1-x)}O_{5x}Cl$ ($0 \leq x \leq 1$) with various S/O ratios, in which a series of low-energy structural configurations can be identified by global energy minimization at each composition between the terminal states. From the convex hull, we can find two stable compounds of $Li_6PS_{4.25}O_{0.75}Cl$ (on the S-rich side, Figure 1b) and $Li_6PSO_4Cl$ (on the O-rich side, Figure 1d) falling on the bottom edge of the energy hull. A slightly metastable compound of $Li_6PS_4OCl$ (Figure 1c marked by a yellow star) is close to the state of $Li_6PS_{4.25}O_{0.75}Cl$, with a slightly higher O concentration. There are tiny structural differences between $Li_6PS_{4.25}O_{0.75}Cl$ and $Li_6PS_4OCl$. In the configuration of $Li_6PS_{4.25}O_{0.75}Cl$, the conventional cell is constructed by three tetrahedral $PS_3O$ units, one tetrahedral $PS_4$ unit, four octahedral $Li_6S$ units, and four $Cl^-$ anions located between the polyhedral units. In the configuration of $Li_6PS_4OCl$, all tetrahedral units are composed by $PS_3O$. It is shocking to note that thermodynamically the O and S mixing in the chalcogen sites of the $Li_6PS_{5(1-x)}O_{5x}Cl$ is not unlimited even by iso-valent substitution on the same sublattice sites. In the range of $0.2<x<0.8$, the energies of mixed states are above the bottom edge of the hull defined by the lowest common tangents, so that these compounds are only metastable and they tend to decompose into the stable phases over equilibration. The equilibrium limits for S/O ratios are thus defined on the chalcogen sites, so that only about 1/5 of S sites can be replaced by O and vice versa. Such thermodynamic characteristics defines the energetic driver for the well- recognized spinodal decomposition in solid state phase transformations, so that coherent structures of the same structural scaffold but different compositions can be intermingled in the nanoscale through equilibration.

In order to maintain the superb-ionic conductivity as the pristine $Li_6PS_5Cl$, we mainly focus on the S-rich compounds of $Li_6PS_{4.25}O_{0.75}Cl$ (Figure 1b) and $Li_6PS_4OCl$ (Figure 1c), since the electronegativity of the chalcogen species needs to be limited to guarantee good ionic conductivity.[9, 15, 35] Besides, it was proven experimentally that a slight off-stoichiometry with more Li and less Cl has remarkable benefit on $Li^+$ ion conductivities, without imbalance of charge neutrality.[9, 44, 53] Thus, the slight off-stoichiometric compositions of $Li_{6.25}PS_{4.2}O_{1.05}Cl_{0.75}$, $Li_{6.25}PS_4O_{1.25}Cl_{0.75}$, and $Li_{6.25}PS_{3.5}O_{1.75}Cl_{0.75}$ were taken forward for experimental assessment in this work.

XRD patterns from as-synthesized $Li_{6.25}PS_{4.2}O_{1.05}Cl_{0.75}$, $Li_{6.25}PS_4O_{1.25}Cl_{0.75}$, and $Li_{6.25}PS_{3.5}O_{1.75}Cl_{0.75}$ samples are presented in Figure 1e. The patterns of $Li_{6.25}PS_{4.2}O_{1.05}Cl_{0.75}$ and $Li_{6.25}PS_4O_{1.25}Cl_{0.75}$ agree well with the XRD pattern of $Li_6PS_5Cl$, with the typical space group symmetry of $F4\bar{3}m(216)$ for the cubic argyrodite phase. Minor impurity peaks of $Li_2S$ appear in the $Li_{6.25}PS_{3.5}O_{1.75}Cl_{0.75}$ sample, suggesting that the O content in this compound is beyond the equilibrium limit. As shown in Figure 1f, the diffraction peaks from O substituted compound shift towards higher diffraction angles, suggesting slightly shrunk crystal cell, due to stronger binding among opposite ions, in line with unit cell sizes in the following order: $Li_6PS_5Cl$ (9.859Å) > $Li_{6.25}PS_{4.2}O_{1.05}Cl_{0.75}$ (9.851Å) > $Li_{6.25}PS_4O_{1.25}Cl_{0.75}$ (9.84Å). Associated lattice information is listed in Table S1.

Raman spectra of the pristine $Li_6PS_5Cl$ (dark green) and $Li_{6.25}PS_4O_{1.25}Cl_{0.75}$ (orange) are shown in Figure 1g. The peaks at 196, 268, 423, 574, and 602 $cm^{-1}$ are attributed to vibrational modes of $PS_4^{3-}$ within $Li_6PS_5Cl$, being consistent with reported values.[9, 54] The Raman pattern of $Li_{6.25}PS_4O_{1.25}Cl_{0.75}$ is similar to that of $Li_6PS_5Cl$. The corresponding peaks for $Li_{6.25}PS_4O_{1.25}Cl_{0.75}$ located at 203.3, 273.7, 430.7, 581.4, and 609.1 $cm^{-1}$. A moderate Raman upshift of 7.7 $cm^{-1}$ at the main peak with respect to that of the pristine $Li_6PS_5Cl$ can be observed in Figure 1h, and there is no Raman evidence of any impurity phases, being consistent with the XRD result.

## 2.2 Ionic Conductivity and Activation Barriers

The impedance spectroscopy of $Li_6PS_5Cl$ and $Li_{6.25}PS_4O_{1.25}Cl_{0.75}$ over a broad frequency range from 10 Hz to 20 MHz at various temperatures from around room temperature (35 °C) to 95 °C are shown in **Figure 2**a and b, with insets of equivalent circuits from the ZView fitting. The impedance spectra for both SSEs are linear without any double-layer resistance in Nyquist diagrams. The ionic conductivity for $Li_6PS_5Cl$ is calculated to be 3.26 mS $cm^{-1}$ at 35 °C. The $Li^+$ ion conductivity of $Li_{6.25}PS_4O_{1.25}Cl_{0.75}$ can still reach 2.8 mS $cm^{-1}$ at 35 °C, though the O substitution induces somewhat stronger binding. The double layer impedance for $Li_6PS_5Cl$ and $Li_{6.25}PS_4O_{1.25}Cl_{0.75}$ only appear at sub-zero Celsius temperatures, when the ionic conductivity becomes mildly poorer. The charge-transfer resistance due to the double layer for $Li_{6.25}PS_4O_{1.25}Cl_{0.75}$ (308 Ω, Figure 2c) is in the same order of that of $Li_6PS_5Cl$ (120 Ω, Figure 2d) at -20 °C. The ionic conductivity of $Li_{6.25}PS_4O_{1.25}Cl_{0.75}$ at -20 °C is 1.49 mS $cm^{-1}$, which is comparable with that of the pristine $Li_6PS_5Cl$ sulfide (2.32 mS $cm^{-1}$). Such ionic conductivity at rather low temperatures in the O-containing SSE is still above the practical request of 1 mS $cm^{-1}$ for good electrolytes, suggesting suitability in application to practical solid battery cells.

The Log (σT) *vs.* (1000/T) plot of the ionic conductivity measured at various temperatures are displayed in Figure 2e. The theoretical ionic conductivity derived from ab initio molecular dynamics (AIMD) simulations (short dash lines) are also shown in Figure 2e as theoretical references from perfect single crystals. It is highly encouraging to note the excellent agreement between the experimental and theoretical data above the room temperature, when no double space charge layers are present. Taking the temperature at 35 °C for example, the theoretically predicted $Li^+$ ion conductivities for $Li_6PS_5Cl$ and $Li_{6.25}PS_4O_{1.25}Cl_{0.75}$ are 3.23 and 2.85 mS $cm^{-1}$, respectively, which agree very well with the corresponding experimental data (3.26 and 2.8 mS $cm^{-1}$). The agreement at higher temperatures is even better.

Below the room temperature, the presence of the space charge layers is associated with lowered conductivity than the theoretically assessed data from perfect single crystals.[55] It is therefore reasonable to attribute such lowered conductivity and associated space charge layers to more evident hindrance to ionic transportation below room temperature, when the particle boundary resistance could get evident. This deviation from the theoretical reference at low temperatures resulted in a slightly higher average activation barriers $E_a$ in experimental tests ($Li_6PS_5Cl$, $E_a$=0.25 eV; $Li_{6.25}PS_4O_{1.25}Cl_{0.75}$, $E_a$=0.33 eV) than the theoretical predictions ($Li_6PS_5Cl$, $E_a$=0.21 eV; $Li_{6.25}PS_4O_{1.25}Cl_{0.75}$, $E_a$=0.27 eV), with the extra activation barrier being attributable to the particle boundaries. The probability density distribution of $Li^+$ ions from the AIMD simulations at the temperature of 1000 K are also indicative of the similar ionic transport performance of $Li_6PS_5Cl$ (Figure 2f) and $Li_{6.25}PS_4O_{1.25}Cl_{0.75}$ (Figure 2g).

### 2.3 Stability against Air and Moisture

Stabilities of $Li_{6.25}PS_4O_{1.25}Cl_{0.75}$ and $Li_6PS_5Cl$ against dry air were evaluated by differential scanning calorimetry (DSC) in an atmosphere with continuous flow 20% $O_2$ and 80% $N_2$.[56-57] From the DSC results in **Figure 3a**, exothermic peaks were observed at a quite low temperature of 75 °C for the $Li_6PS_5Cl$ sample. Exothermic phenomenon occurred drastically for $Li_6PS_5Cl$ in the range from 200 to 300 °C, suggesting that radical reactions between $Li_6PS_5Cl$ and $O_2$ took place. In contrast, for the $Li_{6.25}PS_4O_{1.25}Cl_{0.75}$ sample, no evident exothermal peaks were detected at temperatures below 300 °C, indicating that the $Li_{6.25}PS_4O_{1.25}Cl_{0.75}$ was prominently resistive to dry air. Such remarkable hindrance against reaction with oxygen is in line with the ATAT outcome that further substitution of S by O beyond the 1/5 limit is not thermodynamically favored.

The stability of $Li_6PS_5Cl$ and $Li_{6.25}PS_4O_{1.25}Cl_{0.75}$ electrolyte against humid air was further examined. After being exposed to air with humidity of 53% for 0.5 h, the surface of $Li_6PS_5Cl$ was covered by dense water drops, but no visible water drops were evident on the surface of $Li_{6.25}PS_4O_{1.25}Cl_{0.75}$. XRD patterns of $Li_6PS_5Cl$ and $Li_{6.25}PS_4O_{1.25}Cl_{0.75}$ after exposed to humid air for 0.5h are shown in Figure 3b. The main characteristics of the argyrodite phase was well maintained in the XRD patterns of $Li_{6.25}PS_4O_{1.25}Cl_{0.75}$, while the XRD pattern of $Li_6PS_5Cl$ exhibited numerous peaks from products of yet unknown structures. This demonstrated that the $Li_{6.25}PS_4O_{1.25}Cl_{0.75}$ electrolyte had fairly good resistance to humid air. The samples exposed to humid air were then subjected to 5h annealing at 180 °C under an argon atmosphere. The XRD pattern of the post-annealed $Li_{6.25}PS_4O_{1.25}Cl_{0.75}$ and $Li_6PS_5Cl$ are shown in Figure 3c. It is seen that the $Li_{6.25}PS_4O_{1.25}Cl_{0.75}$ electrolyte only showed rather insignificant evidence from $Li_2S$, with major phase being the argyrodite phase. On the contrary, $Li_2S$ became the main component in the post-annealed $Li_6PS_5Cl$ electrolyte, with the argyrodite phase being almost completely eliminated.

In terms of room-temperature impedances as shown in Figure 3d, a huge double-layer impedance was present in the post-annealed $Li_6PS_5Cl$ exposed to humid air, indicating a sharp decrease of the ionic conductivity of the $Li_6PS_5Cl$. It was amazing to observe that the $Li^+$ ion conductivity of the post-annealed $Li_{6.25}PS_4O_{1.25}Cl_{0.75}$ was still 1.5 mS $cm^{-1}$ at room temperature. This is strong evidence that the $Li_{6.25}PS_4O_{1.25}Cl_{0.75}$ electrolyte is remarkably resilient to humid air. The fact that the $Li_{6.25}PS_4O_{1.25}Cl_{0.75}$ electrolyte is so endurable against humid air is highly desirable for a solid electrolyte,[57-58] so that it can be handled conveniently without stringent control of the processing atmosphere, thus leading to significant reduction of processing cost.

## 2.4 Compatibility with Li Anode

The stability with Li anode for Li|$Li_6PS_5Cl$ and Li|$Li_{6.25}PS_4O_{1.25}Cl_{0.75}$ were compared by galvanostatic cycling tests over the symmetric Li|SSE|Li cells. As displayed in Figure. S1, the total resistance for Li|$Li_6PS_5Cl$|Li before plating and stripping of lithium was only round 120 Ω, which indicates good wetting between $Li_6PS_5Cl$ and Li anode. At a low current density of 0.1 mA $cm^{-2}$, the resistance dropped significantly to around 50 Ω, **Figure S1**. The cycling result for this symmetric cell of Li|$Li_6PS_5Cl$|Li is shown in **Figure 4**a (in green color). As the current density was increased to 0.15 mA $cm^{-2}$ with a cut-off capacity of 0.15 mAh $cm^{-2}$ (Figure 4a), short-circuiting occurred in the Li|$Li_6PS_5Cl$|Li cell during a short cycling time around only 10 h, when the internal cell resistance disappeared, leading to a chaotic impedance spectrum due to shorting. One can see from the SEM images in **Figure S**2a - c and EDS mapping in Figure S2d - f that coarse and densely populated lithium dendrites appeared and grew well into the $Li_6PS_5Cl$ disc over cycling. Besides, the growth of lithium dendrites

tended to induce significant internal stress to generate cracks in the $Li_6PS_5Cl$ disc (Figure S2a and b). The cross-section view of the post-cycling $Li_6PS_5Cl$ disc (**Figure S3**a - e) show lithium dendrites penetrating the disc thus finally resulting in short circuit of the symmetric cell.

For Li|$Li_{6.25}PS_4O_{1.25}Cl_{0.75}$, the initial total resistance was slightly larger than that of Li|$Li_6PS_5Cl$|Li, and it was maintained around 100 Ω without short circuiting even during cycling at a rather high current density of 1 mA $cm^{-2}$ (Figure S1b). The symmetric cell using $Li_{6.25}PS_4O_{1.25}Cl_{0.75}$ as electrolyte exhibited a very stable Li plating and stripping performance, which was able to withstand high current density of 1 mA $cm^{-2}$ as shown in Figure 4b. As is shown in Figure 4c, the cycling at a high current density of 1 mA/$cm^2$ was stable even over an additional 1400 h, with the magnified regions of the stable voltage profile at different time shown as insets in Figure 4c. From the SEM image in Figure S4, one can see that a clean and flat surface of $Li_{6.25}PS_4O_{1.25}Cl_{0.75}$ was maintained, indicating that lithium dendrites were completely suppressed.

The performance of repeated stripping/plating of $Li^+$ in a symmetric cell depends largely on the interfacial stability. The AEI generated on the surface of Li anodes were collected through tape stripping using adhesive tapes. **Figure 5**a and b show the Raman spectrum of the collected SEI samples at the interface of Li|$Li_6PS_5Cl$ and Li|$Li_{6.25}PS_4O_{1.25}Cl_{0.75}$, respectively. In addition to the Raman peaks belonging to $Li_6PS_5Cl$ and $Li_{6.25}PS_4O_{1.25}Cl_{0.75}$, there emerges a new $Li_2S$ peak around 368 $cm^{-1}$[54] in both samples after cycling. The $Li_2S$ peak in $Li_{6.25}PS_4O_{1.25}Cl_{0.75}$ is evidently weaker than that of $Li_6PS_5Cl$ sample (Figure 5a and b). Furthermore, another new peak around 940 $cm^{-1}$, corresponding to the vibrational modes of $PO_4^{3-}$ [59] (consistent with the main peak of $Li_3PO_4$ sample displayed in Figure S5), can be observed in the sample from $Li_{6.25}PS_4O_{1.25}Cl_{0.75}$. The compositions of the SEI identified from Raman spectroscopy are consistent with the DFT simulation results for phase equilibrium at the interface of Li|SSE, with stable phases being listed in Table S2 and S3. The outstanding

difference between the AEIs of the two electrolytes is that the AEI for the oxygen-containing electrolyte has the $Li_3PO_4$ phase as an important constituent.

As is schematically shown in Figure 5c, the $Li_6PS_5Cl$ electrolyte tends to react with the lithium anode to form a thicker but looser AEI layer at the interface.[60-62] Such an AEI layer is not adequately insulating to electrons,[62] which results in the gathering of electrons at cusps and leads to local enhancement of the Li-ion flux.[22] This helps to induce inhomogeneous deposition of Li and finally formation of Li dendrites.[63] Consequently, numerous microscopic cracks (as shown in Figure S2) will be generated in the SSE due to the internal stress caused by the growth of Li dendrites during repeated stripping/plating of $Li^+$ over prolonged cycling.[63] This will in turn make the deposition of Li even more inhomogeneous thus accelerating the evolution of Li dendrites.

In contrast, as shown in Figure 5d, the $Li_3PO_4$ with a very large bandgap of 7.4 eV (Figure. S7 e, is very insulating to electrons, so that its presence in the AEI layer of $Li_{6.25}PS_4O_{1.25}Cl_{0.75}$|Li help to overcome the aforementioned shortcoming, leading to sustained structural integration over the $Li_{6.25}PS_4O_{1.25}Cl_{0.75}$|Li interface owing to effective suppression of lithium dendrites. It is noticed in latest studies,[22, 46, 64] that $Li_3PO_4$ was incorporated in artificial coatings to protect the lithium anode from evolution of lithium dendrite. Here in this work, one can see that in situ formation of $Li_3PO_4$ in the AEI at the Li|$Li_{6.25}PS_4O_{1.25}Cl_{0.75}$ interface was behind the excellent cycling performance of Li|$Li_{6.25}PS_4O_{1.25}Cl_{0.75}$|Li. It is worth mentioning that such *in situ* formed AEI containing the $Li_3PO_4$ does not involve complex and expensive coating technologies, together with more intimate interfacial bonding highly advantageous to electrochemical performance.

## 2.5 The Oxidation Potential

Theoretical electrochemical windows for $Li_6PS_5Cl$ and $Li_6PS_4OCl$ were calculated based on cohesive energies of each low-energy structure on the convex hull against change of lithium content, by global energy minimization through ATAT simulation. The formation energy of ATAT identified structures are referred to that of the stable constituent phases. The resultant voltage profiles are displayed in **Figure 6**a and b. The formation energies of compounds for phase transitions of interest owing the changes in lithium contents are listed in Table S2 and Table S3. Although the $Li_6PS_5Cl$ (with stable constituent phases of $Li_3PS_4$+$Li_2S$+LiCl) and $Li_6PS_4OCl$ (with stable constituent phases of 3/4$Li_3PS_4$+LiCl+$Li_2S$+1/4$Li_3PO_4$) are metastable states with respect to associated constituent phases, the small positive formation energy of 0.0761 and 0.0634 eV/atom can be readily compensated by the entropy terms, so that they can be readily synthesized in practice. The de-lithiation states $Li_2PS_5Cl$ and $Li_2PS_4OCl$ with the depletion of four $Li^+$ from the pristine states, have even smaller formation energies of 0.042 and 0.0433 eV/atom. It is envisaged that loss of more than four $Li^+$ ions could be rather difficult due to considerable positive formation energy beyond 0.1 eV/atom. The maximum oxidation potentials of $Li_6PS_5Cl$ and $Li_{6.25}PS_4O_{1.25}Cl_{0.75}$ (with respect to the maximum de-lithiation states) can be determined to be 2.19-2.93 and 2.49-3.33 V, respectively, using equation (1).

Cyclic voltammetry (CV) tests of cells in the form of Li|SSE|SSE+C were carried out, with the SSE+C counter electrode (with 25 wt% carbon) being used to promote the redox kinetics for better accuracy in measuring the oxidation potentials.[12, 65] As presented in Figure 6c, the oxidation peak for $Li_6PS_5Cl$ emerged at 2.54 V and the one for $Li_{6.25}PS_4O_{1.25}Cl_{0.75}$ occurred at 3.55 V. The onset of oxidation for $Li_6PS_5Cl$ was 2.22 V and that for $Li_{6.25}PS_4O_{1.25}Cl_{0.75}$ was 3.1 V. It is amazing to note that the oxidation potential was elevated for around 1 V by substituting a fifth of S with O in the $Li_6PS_5Cl$ electrolyte. One notes in Figure 6 that the experimentally measured oxidation potentials of both $Li_6PS_5Cl$ and $Li_{6.25}PS_4O_{1.25}Cl_{0.75}$ are in the theoretically predicted ranges of 2.19-2.93 and 2.49-3.33 V, respectively.

### 2.6 Compatibility with bare $LiCoO_2$ Cathode

Owing to the remarkable elevation of the oxidation potential, the $Li_{6.25}PS_4O_{1.25}Cl_{0.75}$ is expected to tolerate high voltage cathode materials. Type-2032 coin cells were fabricated through simple cold pressing, as shown in **Figure 7**a. The SSE disc of $Li_6PS_5Cl$ or $Li_{6.25}PS_4O_{1.25}Cl_{0.75}$ was introduced between the Li anode and composite cathode disc made of bare $LiCoO_2$ powder, a SSE, and carbon black. Galvanostatic charge–discharge tests were performed on the full battery cells in a voltage window of 2.5–4.2 V (vs. $Li/Li^+$). Figure 7b shows the charge and discharge voltage profiles of the $LiCoO_2|Li_6PS_5Cl|Li$ and $LiCoO_2|Li_{6.25}PS_4O_{1.25}Cl_{0.75}|Li$ cell at 0.05 mA $cm^{-2}$ (with a $LiCoO_2$ mass loading of 4 mg $cm^{-2}$) during the 1st, 2nd, and 10th cycling. The $LiCoO_2|Li_6PS_5Cl|Li$ cell showed an initial charge capacity of 152.6 mAh $g^{-1}$ and a discharge capacity of 80 mAh $g^{-1}$, corresponding to an initial Coulombic efficiency of only 52.4%, which is attributable to the significant interfacial reaction at the anode-SS interface with decomposition of $Li_6PS_5Cl$.[19] The capacity of the $LiCoO_2|Li_6PS_5Cl|Li$ cell degraded rapidly, dropped to 68.6 mAh $g^{-1}$ after only 45 cycles beyond which the cell failed to operate, Figure 7c. As is compared in Figure 7, the $LiCoO_2|Li_{6.25}PS_4O_{1.25}Cl_{0.75}|Li$ cell showed an initial charge capacity of 151.9 mAh $g^{-1}$ and a discharge capacity of 130.8 mAh $g^{-1}$. The initial Coulombic efficiency was significantly increased up to 86%, which immediately increased to over 99.6% after only several initial cycles. The $LiCoO_2|Li_{6.25}PS_4O_{1.25}Cl_{0.75}|Li$ cell then exhibited a stable steady-state evolution of capacity, with over 90.8 mAh $g^{-1}$ still being maintained after 100 cycles.

In order to find out what happened at the electrolyte-anode interface over charging-discharging cycling, the full cells $LiCoO_2|Li_6PS_5Cl|Li$ and $LiCoO_2|Li_{6.25}PS_4O_{1.25}Cl_{0.75}|Li$ were disassembled after 10 cycles charged to 4.2 V. The CEI formed between $LiCoO_2$ and the two SSEs in the composite cathode were characterized by Raman tests, shown in Figure 7d and e.

At the interface of $LiCoO_2|Li_6PS_5Cl$, an obvious peak belonging to $Li_2S$ is evident in Figure 7d. At the interface of $LiCoO_2|Li_{6.25}PS_4O_{1.25}Cl_{0.75}$, the chemical components of $PO_4^{3-}$ is present in the Raman spectrum in Figure 7e, albeit without evidence of the $Li_2S$ phase. The finding of $PO_4^{3-}$ is consistent with the DFT modeling result that $Li_3PO_4$ tends to form above 3.33 V, Table S3.

XPS spectra from the composite cathodes before and after cycling are displayed in **Figure 8**. The S 2p3/2 spectrum of the cathode ($LiCoO_2$+$Li_6PS_5Cl$) before cycling is shown in Figure 8a. It can be deconvoluted into three components. The main doublet of S 2p3/2 at 160.78 eV (red component) is attributed to the sulfur in $PS_4^{3-}$.[19, 36] The doublet at 159.58 eV (magenta component) can be attributed to the sulfur in $Li_2S$, which corresponds to a stable constituent of CEI between $LiCoO_2$ and $Li_6PS_5Cl$.[19] The final component at 162.38 eV (blue) can be assigned to the sulfur in polysulfide $P_2S_x$.[19] The P 2p3/2 spectrum from the composite cathode, Figure 8b, consists of two components. The doublet at a 2p3/2 with a binding energy of 131.08 eV (orange component) is the fingerprint peak of P in the argyrodite structure.[19, 27] The existence of $P_2S_x$ is also confirmed by the P 2p doublet at 132.98 eV (dark yellow component).[19] The detection of $Li_2S$ and $P_2S_x$ suggests that reaction happened at $LiCoO_2|Li_6PS_5Cl$ interface in the as-prepared composite cathode even before it was subjected to charging or discharging.[19]

After 10 cycles charged to 4.2 V, the intensity of both the S 2p peaks belong to $Li_xS$ and $P_2S_x$ increased (Figure 8c). Besides, an additional component at around 166.28 eV can be attributed to trace sulfate ($SO_3^{2-}$)[19] from possible reaction with the oxide cathode at $LiCoO_2|Li_6PS_5Cl$. For the P 2p spectra (Figure 8d), a new peak corresponding to $P_2S_5$ at 133.68 eV emerges (pink), which indicates that a $Li^+$ depletion layer may be generated at $LiCoO_2|Li_6PS_5Cl$ after 10 cycles. This is consistent with DFT results listed in Table S2, which predicts that $LiS_4$ and $P_2S_5$ tend to be generated at high voltage.

On the other hand, for the composite cathode of $LiCoO_2$+$Li_{6.25}PS_4O_{1.25}Cl_{0.75}$ before charging/discharging, the S 2p spectrum, Figure 8e, also shows a main doublet at 160.88 eV (red component) from the sulfur in $Li_{6.25}PS_4O_{1.25}Cl_{0.75}$, but there is no evidence for the presence of $P_2S_x$ and the $Li_2S$ peak is trivial. It is also noted that only one doublet at 133.08 eV corresponds to the P 2p in $PS_3O^{3-}$ (Figure 8f), but without any impurity peaks from $P_2S_x$. This suggests that the interface stability between $LiCoO_2$ and $Li_{6.25}PS_4O_{1.25}Cl_{0.75}$ is largely improved owing to the incorporation of oxygen in $Li_{6.25}PS_4O_{1.25}Cl_{0.75}$.

After 10 cycles charged to 4.2 V, the S 2p from $SO_3^{2-}$ and $P_2S_5$ are still invisible in the XPS spectrum, Figure 8g. The P 2p spectrum is shifted to lower binding energy after cycling (Figure 8h), which can be fitted into three components. The main component at 131.28 eV (violet component) is assigned to $PO_4^{3-}$, according to the P 2p spectrum from $Li_3PO_4$ as reference standard, shown in Figure 8i. Trace amount of $P_2S_5$ is also detected as the minor decomposition product of the electrolyte. The XPS findings about the generation of $Li_3PO_4$ and the resilience against decomposition of $Li_{6.25}PS_4O_{1.25}Cl_{0.75}$ is consistent with the aforementioned Raman result.

Schematic diagrams for the interphases generated at the interface of $LiCoO_2|Li_6PS_5Cl$ and $LiCoO_2|Li_{6.25}PS_4O_{1.25}Cl_{0.75}$ are shown in Figure 8j and k, respectively. At the interface of $LiCoO_2|Li_6PS_5Cl$, a thick Li depletion layer was developed, since the $Li_6PS_5Cl$ is less resistive to voltage rise at the cathode-electrolyte interface during charging, which led to significant decomposition of the SSE into a mixture of binary compounds with poor ionic conductivity, thus leading to significantly slowed kinetic processes over charging-discharging and degradation of cell performance. In contrast, since the oxygen-containing SSE is significantly more resilient to voltage rise, and $Li_3PO_4$ was the major product at higher voltage over early cycles. Such a lithium phosphate was usually employed as an effective buffer layer to protect high-voltage oxide cathode materials.[66] Fundamentally, this is attributable to its high oxidation potential of 4.552 V (as shown in Figure S6, from ATAT modeling). The

electrochemical window of $Li_3PO_4$ is from 0.7 ~ 4.552 V, and its presence at the cathode-electrolyte interface provides an ideal *in situ* buffer to protect both the cathode and the SSE from decomposition at elevated voltage, owing to overlapped electrochemical windows at the interface.[35, 44]

**2.7 Electronic Conductivity**

The electronic conductivities of $Li_6PS_5Cl$ and $Li_{6.25}PS_4O_{1.25}Cl_{0.75}$ were measured by the direct current (DC) polarization method. The testing voltage was set from 1 V to 2 V. As shown in **Figure S7**a, the measured electronic conductivity of $Li_6PS_5Cl$ is $5.75\times10^{-8}$ S/cm. The electronic conductivity of $Li_{6.25}PS_4O_{1.25}Cl_{0.75}$ is $1.94\times10^{-8}$ S/cm, which is only about third of that of $Li_6PS_5Cl$. From the electronic band structures of $Li_6PS_5Cl$ and $Li_{6.25}PS_4O_{1.25}Cl_{0.75}$ displayed in Figure S7b and c, both can be classified into intrinsic wide-gap semiconductors with similar band gaps, so that they are naturally poor conductors to electrons. For a direct comparison of conduction band minimum (CBM) (Figure S7d), CBM of $Li_{6.25}PS_4O_{1.25}Cl_{0.75}$ is flatter than that of $Li_6PS_5Cl$, implying that the electronic insulation of $Li_{6.25}PS_4O_{1.25}Cl_{0.75}$ is naturally better.[67-68] Besides, in a full solid cell, the *in-situ* electronic insulator of $Li_3PO_4$ at the interface of Li|$Li_{6.25}PS_4O_{1.25}Cl_{0.75}$ will provides further assistance against self-discharging during service.

**3. Conclusion**

Substitution of S by O in the Argyrodite $Li_6PS_5Cl$ solid electrolyte has been realized under the guidance of first-principles modeling, leading to the discovery of $Li_{6.25}PS_4O_{1.25}Cl_{0.75}$ as a highly promising solid electrolyte which has a superb ionic conductivity of 1.49 mS $cm^{-1}$ even at -20 ℃ with significantly enhanced durance to voltage up to 3.3 V. It is appealing that such an ionic conductor has marvelous compatibility with both the high voltage cathode such as $LiCoO_2$ and the pure lithium anode, with *in situ* $Li_3PO_4$ in AEI and CEI being highly protective of both the anode and the electrolyte from undesirable interfacial reactions. The *in situ* AEI enabled ultra-stable Li plating/stripping at a high current density up to 1 mAh cm$^{-2}$ in Li|$Li_{6.25}PS_4O_{1.25}Cl_{0.75}$|Li symmetric cells over 1800 hours, and the joint benefit from both electrolyte-electrode interphases facilitated the $LiCoO_2$|$Li_{6.25}PS_4O_{1.25}Cl_{0.75}$|Li battery cell to achieve markedly improved cycling stability at a high areal loading of the active cathode material (4 mg cm$^{-2}$).

The high materials stability of the new $Li_{6.25}PS_4O_{1.25}Cl_{0.75}$ electrolyte lies in the thermodynamic characteristics about mixing between S and O on the chalcogen sublattice, with a fundamental spinodal gap being defined by about 1/5 mixing from either minor species, i.e. between $Li_6S_4OCl$ and $Li_6SO_4Cl$. Slight excess in Li and O with suitable reduction of Cl to maintain the overall charge neutrality was effective in enhancing ionic conductivity.

The thermodynamic limitation on O substitution of S leads to significant resilience against oxygen and even humid air. There was slight change in the ionic conductivity of the post-annealed $Li_{6.25}PS_4O_{1.25}Cl_{0.75}$ sample after exposure to humid air (53% humidity). This work offers a highly attractive solid electrolyte in the argyrodite sulfide family, so that high-performance solid battery cells could even be fabricated in ambient air.

**4. Methods**

*Raw materials:* All materials were used directly without any purification, such as $Li_2O$, LiCl, $P_2O_5$, and $P_2S_5$ (all from Aladdin, 99.99%). And $LiCoO_2$ was purchased from Sigma-Aldrich. $Li_2S$ was synthesized using the mechanical milling method reported in our recent work.[9]

*Materials synthesis:* Ball milling towards the mixed powders of $Li_2O$, $Li_2S$, $P_2O_5$, $P_2S_5$, and LiCl with designated molar ratios was carried out at an initial low speed of 250 rpm for 2 h, followed by high-speed ball milling at 550 rpm for another 12 h to obtain the precursor. The ball-milled precursor was then cold pressed into discs (with thickness about 0.5 mm; diameter of 13 mm) under 300Mpa. The discs were sealed in quartz tubes, followed by annealing at 530 °C for 10 h in a muffle furnace and then cooled down to room temperature at 1 °C per minute to complete the synthesis of each SSE sample. All of the processes were carried out in a glovebox under an Ar atmosphere to isolate the materials from humidity and air.

*Phase identification and materials characterization:* X-ray diffraction (XRD, Rigaku Ultima IV) with Cu-Ka radiation was employed for the phase identification, when the samples were sealed within polyimide envelops.

The microstructures and morphologies were characterized by field emission scanning electron microscopy (SEM, ZEISS, SIGMA 500/VP), equipped with an energy dispersive X-ray spectrometer (EDS) for elemental analysis.

Raman spectroscopy was performed using a LabRAM HR Evolution facility with He/Ne laser excitation at 532 nm and a grating of 600 lines per mm. The samples were sealed in the glass holder by polyimide film.

X-ray photoelectron spectroscopy (XPS) was carried out with a Krotos AXIS Ultra Spectrometer system using a monochromatic Al K(alpha) source (25 mA, 15 kV).

*Ionic conductivity measurements:* Each SSE disc sample was put into a copper clapping assembly within an Ar filled glovebox, which helped secure good contact to the electrodes and isolate the sample from air. The clapped sample was then subject to each specified temperature in a dry pre-heated chamber, keeping for 0.5 h to settle the sample temperature before impedance measurements.

Ionic conductivity was measured by electrochemical impedance spectroscopy (EIS) with an applied frequency from 10 Hz to 20 MHz, using a Schlumberger Solartron 1260 frequency response analyzer at a sinusoidal amplitude of 10 mV.

The ionic conductivity can then be derived as, $\sigma_{total} = \frac{d}{A * R_{total}}$, where $d$ is the disc thickness, $A$ is the area of disc, and $R_{total}$ is the total resistance from the test of EIS.

*Air stability assessment:* The electrolyte stability against dry air was assessed by differential scanning calorimetry (DSC). Approximately 15 mg of the powder sample was loaded into an alumina sample holder in the glovebox. DSC was carried out with the sample under a continuous flow of mixed 20% $O_2$ and 80% $N_2$. The temperature range for DSC analysis was set from room temperature to 350 °C at a heating rate of 5 °C $min^{-1}$.

The stability against moisture was evaluated as follows: Electrolytes were placed in the air at humidity of 53%. After being exposed to the humid air for half an hour, XRD was adopted to identify the changes in the phase or purities in the sample.

The sample was then subjected to post-annealing in the muffle furnace (180 °C) for 5 hours, followed then by XRD and the EIS measurements again.

*Cell assembly and electrochemical measurements:* Each SSE disc and full battery cell assemblies were prepared in an Ar-filled glovebox.

The SSE–C|SSE|Li cells for cyclic voltammetry tests were prepared by direct cold-pressing of a piece of Li metal foil (China Energy Lithium Co., Ltd.) as anode, a layer of 100 mg SSE powder as solid electrolyte, and a layer of 10 mg SSE+C (with a weight rate of SSE:C=75:25) powder as cathode under 300 MPa. The cyclic voltammetry tests were carried out in the voltage range from 0 to 5 V at a scan rate of 0.01 mV $s^{-1}$.

To fabricate the symmetric Li|SSE|Li cells , typically, 100 mg of SSE electrolyte was cold pressed by 300 MPa to form a solid disc. Two pieces of Li metal were placed onto both sides of the SSE disc to assemble a symmetric cell. Galvanostatic cycling over the cells were carried out using a LAND battery testing station (CT-2001A, Wuhan Rambo Testing Equipment Co., Ltd.). The current density and cut-off capacity were set to be from 0.05 mA $cm^{-2}$ (0.05 mAh $cm^{-2}$) to 1 mA $cm^{-2}$ (1 mAh $cm^{-2}$).

ASSBs, in the form of Li|SSE|LCO+C+SSE were fabricated in three steps: (a) Electrolyte powder with around 100 mg was pressed by 100 MPa to form a solid disc with diameter of 13 mm. (b) The composite cathode around 8 mg (with weight rate of LCO:SSE:C =5:4:1) was uniformly spread on one side of solid disc and pressed under ≈300 MPa for 5 min. (c) A piece of Li foil was then attached on the other side of solid disc at a pressure of 120 MPa to form a three-layered disc, which was sandwiched between stainless-steel gaskets as current collectors and finally sealed into coin cells. The voltage window was set as 2.5–4.2 V (*vs.* $Li/Li^+$) to evaluate the cycling stability with the constant current density of 0.05 $mA/cm^2$.

*Electronic conductivity measurements:* The electronic conductivities of $Li_6PS_5Cl$ and $Li_{6.25}PS_4O_{1.25}Cl_{0.75}$ were measured with the Hebb-Wagner polarization method over Ag|SSE|Ag cells, and the results are shown in Figure S2a The total electronic conductivity are obtained from the linear region of Figure S2a using the equation $(\sigma_e + \sigma_h) = (L/S)\ (d|I_{e,h}|\ /dE)$, where $\sigma_e$ and $\sigma_h$ are conductivities of electron and hole, respectively, $L$ and $S$ are the thickness and the area of the solid electrolyte layer, $I_{e,h}$ is the sum of the electron and hole currents, and $E$ is the polarization voltage.

*DFT modeling:* Theoretical calculations were performed in the framework of density functional theory (DFT), using the Vienna Ab Initio Simulation Package (VASP).[69-70] The ionic potentials including the effect of core electrons are described by the projector augmented wave (PAW) method,[71-72] with the Perdew–Burke–Ernzerhof (PBE) GGA exchange–correlation (XC) functional being used to account for the valence electrons for structure energy calculations.[73-74] The convergence criteria for the relaxation and the associated input parameters were based on systematic tests in the team.[75-77] An actual grid spacing smaller than 0.04 $Å^{-1}$ was chosen as a standard to divide the reciprocal space, with the cutoff energy being set at 500 eV. All structures were geometrically relaxed until the total force on each ion is reduced below 0.01 eV $Å^{-1}$.

The Alloy-Theoretic Automated Toolkit (ATAT)[78-79] was used to identify the most stable phase for each composition with different lithium contents while keeping the amounts of the other elements unchanged, with global energy minimization assessed using total energies for various potential phase structures generated by the linked VASP calculation.

The average electrochemical potential $\boldsymbol{\bar{V}_{A\rightarrow B}}$, for the transition between state A ($\boldsymbol{Li_x}\prod$) and state B ($\boldsymbol{Li_{x+\Delta x}}\prod$), with reference to electrochemical potential v.s. Li/Li$^+$ related to total energies ($E_{total}$) can be calculated as follows:[35, 77]

$$\boldsymbol{\bar{V}_{A\rightarrow B} = -1/z\{[E_{total}(Li_{x+\Delta x}}\prod) - \boldsymbol{E_{total}(Li_x}\prod)]/\Delta\mathbf{x} - \boldsymbol{E_{total}(Li)\}}, \quad (1)$$

where x is the number of Li in the formula unit of $\boldsymbol{Li_x}\prod$, the charge value z=1 for Li$^+$, $\Delta\mathbf{x}$ is the change in the number of Li atoms, and $\prod$ refers to the collection of other constituents.

Ab initio molecular dynamics (AIMD)[77, 80-83] is carried out to study the ionic transport behavior at elevated temperatures, thus providing a statistic insight towards the ionic diffusion.

The ionic conductivity σ is related to the diffusion coefficient in the Arrhenius form of $D = D_0 \exp(-\frac{E_a}{k_B T})$ by the Nernst–Einstein equation,[84]

$$\sigma = \frac{\rho {z_a}^2 F^2}{RT} D = \frac{\rho {z_a}^2 F^2}{RT} D_0 \exp(-\frac{E_a}{k_B T}) = \frac{A}{T} \exp\left(-\frac{E_a}{k_B T}\right) \quad (2)$$

where $\rho$ is the molar density of diffusing alkali metal ions in the unit cell, $z_a$ is the charge number on alkali metal ions (+1 for Li$^+$), *F* and *R* are the Faraday constant and gas constant, respectively, $D_0$ is a constant, $E_a$ is the activation energy for diffusion, and $k_B$ is the Boltzmann constant.

Electronic band structures are calculated using the HSE06[85-87] functional to account for the non-local effect in the XC functionals. We employed a convergence criterion of $10^{-6}$ eV, suitable for electronic self-consistent cycles.

Formation energy ($E_f$) is defined as,

$$E_f = (E_c - E_{eq})/n, \quad (3)$$

in which $E_c$ is total energy of the structural configuration due to change of lithium content in a SSE, $E_{eq}$ is total energy of constituent phases.

**Supporting Information**

Supporting Information is available from the Wiley Online Library or from the author.

**Acknowledgements**

This work is supported in part by the Zhengzhou Materials Genome Institute, the National Natural Science Foundation of China (No. 51001091, 51571182, 111174256, 91233101, 51602094, 11274100, 51571182), the Fundamental Research Program from the Ministry of Science and Technology of China (no. 2014CB931704), and the Program for Science & Technology Innovation Talents in the Universities of Henan Province (18HASTIT009).

**Conflict of Interest**

The authors declare no conflict of interest.

**References**

[1] W. D. Richards, L. J. Miara, Y. Wang, J. C. Kim, G. Ceder, *Chem. Mater.* **2016**, 28, 266.

[2] J. B. Goodenough, K.-S. Park, *J. Am. Chem. Soc.* **2013**, 135, 1167.

[3] Y. Kato, S. Hori, T. Saito, K. Suzuki, M. Hirayama, A. Mitsui, M. Yonemura, H. Iba, R. Kanno, *Nat. Energy* **2016**, 1, 16030.

[4] N. Kamaya, K. Homma, Y. Yamakawa, M. Hirayama, R. Kanno, M. Yonemura, T. Kamiyama, Y. Kato, S. Hama, K. Kawamoto, A. Mitsui, *Nat. Mater.* **2011**, 10, 682.

[5] Y.-S. Hu, *Nat. Energy* **2016**, 1, 16042.

[6] J. C. Bachman, S. Muy, A. Grimaud, H. H. Chang, N. Pour, S. F. Lux, O. Paschos, F. Maglia, S. Lupart, P. Lamp, L. Giordano, Y. Shao-Horn, *Chem. Rev.* **2016**, 116, 140.

[7] H.-J. Deiseroth, S.-T. Kong, H. Eckert, J. Vannahme, C. Reiner, T. Zaiß, M. Schlosser, *Angew. Chem., Int. Ed.* **2008**, 47, 755.

[8] Y. Seino, T. Ota, K. Takada, A. Hayashi, M. Tatsumisago, *Energy Environ. Sci.* **2014**, 7, 627.

[9] M. Xuan, W. Xiao, H. Xu, Y. Shen, Z. Li, S. Zhang, Z. Wang, G. Shao, *J. Mater. Chem. A* **2018**, 6, 19231.

[10] Z. Deng, Z. Wang, I.-H. Chu, J. Luo, S. P. Ong, *J. Electrochem. Soc.* **2015**, 163, A67.

[11] A. Kato, M. Yamamoto, A. Sakuda, A. Hayashi, M. Tatsumisago, *ACS Appl. Energy Mater.* **2018**, 1, 1002.

[12] A. Banerjee, X. Wang, C. Fang, E. A. Wu, Y. S. Meng, *Chem. Rev.* **2020**, 120, 6878.

[13] S. Wenzel, S. Randau, T. Leichtweiß, D. A. Weber, J. Sann, W. G. Zeier, J. R. Janek, *Chem. Mater.* **2016**, 28, 2400.

[14] Y. Xiao, Y. Wang, S.-H. Bo, J. C. Kim, L. J. Miara, G. Ceder, *Nat. Rev. Mater.* **2019**, 5, 105.

[15] Z. Wang, G. Shao, *J. Mater. Chem. A* **2017**, 5, 21846.

[16] D. Lin, Y. Liu, Y. Cui, *Nat. Nanotechnol.* **2017**, 12, 194.

[17] D. Lin, Y. Liu, W. Chen, G. Zhou, K. Liu, B. Dunn, Y. Cui, *Nano Lett.* **2017**, 17, 3731.

[18] Y. Zhu, X. He, Y. Mo, *ACS Appl. Mater. Interfaces* **2015**, 7, 23685.

[19] J. Auvergniot, A. Cassel, J.-B. Ledeuil, V. Viallet, V. Seznec, R. Dedryvère, *Chem. Mater.* **2017**, 29, 3883.

[20] R. Zhang, X. Chen, X. Shen, X.-Q. Zhang, X.-R. Chen, X.-B. Cheng, C. Yan, C.-Z. Zhao, Q. Zhang, *Joule* **2018**, 2, 764.

[21] G. Zheng, S. W. Lee, Z. Liang, H.-W. Lee, K. Yan, H. Yao, H. Wang, W. Li, S. Chu, Y. Cui, *Nat. Nanotechnol.* **2014**, 9, 618.

[22] N. W. Li, Y. X. Yin, C. P. Yang, Y. G. Guo, *Adv. Mater.* **2016**, 28, 1853.

[23] S. Tang, Q. Lan, L. Xu, J. Liang, P. Lou, C. Liu, L. Mai, Y.-C. Cao, S. Cheng, *Nano Energy* **2020**, 71, 104600.

[24] Q. Wang, X. Li, X. He, L. Cao, G. Cao, H. Xu, J. Hu, G. Shao, *J. Power Sources* **2020**, 455, 227988.

[25] C. Wang, H. Xie, L. Zhang, Y. Gong, G. Pastel, J. Dai, B. Liu, E. D. Wachsman, L. Hu, *Adv. Energy Mater.* **2018**, 8, 1701963.

[26] Z. Tu, S. Choudhury, M. J. Zachman, S. Wei, K. Zhang, L. F. Kourkoutis, L. A. Archer, *Nat. Energy* **2018**, 3, 310.

[27] J. Auvergniot, A. Cassel, D. Foix, V. Viallet, V. Seznec, R. Dedryvère, *Solid State Ionics* **2017**, 300, 78.

[28] Y. Li, D. Cao, W. Arnold, Y. Ren, C. Liu, J. B. Jasinski, T. Druffel, Y. Cao, H. Zhu, H. Wang, *Energy Storage Materials* **2020**, 31, 344.

[29] W. Tu, Y. Wen, C. Ye, L. Xing, K. Xu, W. Li, *Energy Environ. Mater.* **2020**, 3, 19.

[30] Y. Ma, *Energy Environ. Mater.* **2018**, 1, 148.

[31] N. Ohta, K. Takada, I. Sakaguchi, L. Zhang, R. Ma, K. Fukuda, M. Osada, T. Sasaki, *Electrochem. Commun.* **2007**, 9, 1486.

[32] G. Oh, M. Hirayama, O. Kwon, K. Suzuki, R. Kanno, *Chem. Mater.* **2016**, 28, 2634.

[33] C. Liu, G. Cao, Z. Wu, J. Hu, H. Wang, G. Shao, *ACS Appl. Mater. Interfaces* **2019**, 11, 31991.

[34] X. Chen, W. He, L.-X. Ding, S. Wang, H. Wang, *Energy Environ. Sci.* **2019**, 12, 938.

[35] H. Xu, Y. Yu, Z. Wang, G. Shao, *J. Mater. Chem. A* **2019**, 7, 5239.

[36] J. Zhang, C. Zheng, L. Li, Y. Xia, H. Huang, Y. Gan, C. Liang, X. He, X. Tao, W. Zhang, *Adv. Energy Mater.* **2019**, 10, 1903311.

[37] K. Takada, N. Ohta, Y. Tateyama, *J. Inorg. Organomet. Polym. Mater.* **2014**, 25, 205.

[38] J. Haruyama, K. Sodeyama, L. Han, K. Takada, Y. Tateyama, *Chem. Mater.* **2014**, 26, 4248.

[39] A. Sakuda, A. Hayashi, M. Tatsumisago, *Chem. Mater.* **2010**, 22, 949.

[40] T. K. Schwietert, V. A. Arszelewska, C. Wang, C. Yu, A. Vasileiadis, N. J. J. de Klerk, J. Hageman, T. Hupfer, I. Kerkamm, Y. Xu, E. van der Maas, E. M. Kelder, S. Ganapathy, M. Wagemaker, *Nat. Mater.* **2020**, 19, 428.

[41] F. Zheng, M. Kotobuki, S. Song, M. O. Lai, L. Lu, *J. Power Sources* **2018**, 389, 198.

[42] Z. Zhang, Y. Shao, B. Lotsch, Y.-S. Hu, H. Li, J. Janek, L. F. Nazar, C.-W. Nan, J. Maier, M. Armand, L. Chen, *Energy Environ. Sci.* **2018**, 11, 1945.

[43] N. Ohta, K. Takada, L. Zhang, R. Ma, M. Osada, T. Sasaki, *Adv. Mater.* **2006**, 18, 2226.

[44] W. Xiao, H. Xu, M. Xuan, Z. Wu, Y. Zhang, X. Zhang, S. Zhang, Y. Shen, G. Shao, *J. Energ. Chem.* **2021**, 53, 147.

[45] Y. Xiao, L. J. Miara, Y. Wang, G. Ceder, *Joule* **2019**, 3, 1252.

[46] B. Huang, X. Yao, Z. Huang, Y. Guan, Y. Jin, X. Xu, *J. Power Sources* **2015**, 284, 206.

[47] J. Y. Liang, X. X. Zeng, X. D. Zhang, P. F. Wang, J. Y. Ma, Y. X. Yin, X. W. Wu, Y. G. Guo, L. J. Wan, *J. Am. Chem. Soc.* **2018**, 140, 6767.

[48] Y. C. Yin, Q. Wang, J. T. Yang, F. Li, G. Zhang, C. H. Jiang, H. S. Mo, J. S. Yao, K. H. Wang, F. Zhou, H. X. Ju, H. B. Yao, *Nat. Commun.* **2020**, 11, 1761.

[49] Y. Zhao, K. Zheng, X. Sun, *Joule* **2018**, 2, 2583.

[50] G. Liu, W. Weng, Z. Zhang, L. Wu, J. Yang, X. Yao, *Nano Lett.* **2020**, DOI: 10.1021/acs.nanolett.0c0248910.1021/acs.nanolett.0c02489.

[51] P. Adeli, J. D. Bazak, K. H. Park, I. Kochetkov, A. Huq, G. R. Goward, L. F. Nazar, *Angew. Chem. Int. Ed.* **2019**, 58, 8681.

[52] S.-T. Kong, H.-J. Deiseroth, J. Maier, V. Nickel, K. Weichert, C. Reiner, *Z. Anorg. Allg. Chem.* **2010**, 636, 1920.

[53] D. Li, L. Cao, C. Liu, G. Cao, J. Hu, J. Chen, G. Shao, *Appl. Surf. Sci.* **2019**, 493, 1326.

[54] Y. Zhou, C. Doerrer, J. Kasemchainan, P. G. Bruce, M. Pasta, L. J. Hardwick, *Batteries & Supercaps* **2020**, 3, 647.

[55] H. Xu, M. Xuan, W. Xiao, Y. Shen, Z. Li, Z. Wang, J. Hu, G. Shao, *ACS Appl. Energy Mater.* **2019**, 2, 6288.

[56] Y. Tian, T. Shi, W. D. Richards, J. Li, J. C. Kim, S.-H. Bo, G. Ceder, *Energy Environ. Sci.* **2017**, 10, 1150.

[57] T. Asano, A. Sakai, S. Ouchi, M. Sakaida, A. Miyazaki, S. Hasegawa, *Adv. Mater.* **2018**, 30, 1803075.

[58] J. Schnell, T. Günther, T. Knoche, C. Vieider, L. Köhler, A. Just, M. Keller, S. Passerini, G. Reinhart, *J. Power Sources* **2018**, 382, 160.

[59] A. Santini, V. Miletic, *J Dent.* **2008**, 36, 683.

[60] T. Liu, L. Lin, X. Bi, L. Tian, K. Yang, J. Liu, M. Li, Z. Chen, J. Lu, K. Amine, K. Xu, F. Pan, *Nat. Nanotechnol.* **2019**, 14, 50.

[61] X.-Q. Zhang, X.-B. Cheng, Q. Zhang, *Adv. Mater. Interfaces* **2018**, 5, 1701097.

[62] Y. Shao, H. Wang, Z. Gong, Dawei Wang, B. Zheng, J. Zhu, Y. Lu, Y.-S. Hu, X. Guo, H. Li, X. Huang, Y. Yang, C.-W. Nan, L. Chen, *ACS Energy Lett.* **2018**, 3, 1212.

[63] J. Kasemchainan, S. Zekoll, D. Spencer Jolly, Z. Ning, G. O. Hartley, J. Marrow, P. G. Bruce, *Nat. Mater.* **2019**, 18, 1105.

[64] T. Deng, X. Ji, Y. Zhao, L. Cao, S. Li, S. Hwang, C. Luo, P. Wang, H. Jia, X. Fan, X. Lu, D. Su, X. Sun, C. Wang, J. G. Zhang, *Adv. Mater.* **2020**, 32, 2000030.

[65] F. Han, Y. Zhu, X. He, Y. Mo, C. Wang, *Adv. Energy Mater.* **2016**, 6, 1501590.

[66] S. Deng, X. Li, Z. Ren, W. Li, J. Luo, J. Liang, J. Liang, M. N. Banis, M. Li, Y. Zhao, X. Li, C. Wang, Y. Sun, Q. Sun, R. Li, Y. Hu, H. Huang, L. Zhang, S. Lu, J. Luo, X. Sun, *Energy Storage Materials* **2020**, 27, 117.

[67] G. Shao, *J. Phys. Chem. C* **2009**, 113, 6800.

[68] G. Shao, *J. Phys. Chem. C* **2008**, 112, 18677.

[69] G. Kresse, J. Hafner, *Phys. Rev. B* **1994**, 49, 14251.

[70] G. Kresse, J. Hafner, *Phys. Rev. B* **1993**, 47, 558.

[71] P. E. Blöchl, *Phys. Rev. B* **1994**, 50, 17953.

[72] G. Kresse, D. Joubert, *Phys. Rev. B* **1999**, 59, 1758.

[73] G. Kresse, J. Hafner, *Phys. Rev. B* **1993**, 48, 13115.

[74] J. P. Perdew, K. Burke, M. Ernzerhof, *Phys. Rev. Lett.* **1996**, 77, 3865.

[75] H. Xu, Y. Yu, Z. Wang, G. Shao, *Energy Environ. Mater.* **2019**, 2, 234.

[76] Z. Wang, M. Deng, X. Xia, Y. Gao, G. Shao, *Energy Environ. Mater.* **2018**, 1, 174.

[77] Z. Wang, H. Xu, M. Xuan, G. Shao, *J. Mater. Chem. A* **2018**, 6, 73.

[78] A. V. D. Walle, G. Ceder, *J. Phase Equilib.* **2020**, 23, 348.

[79] A. V. D. Walle, M. Asta, G. Ceder, *Calphad* **2002**, 26, 539.

[80] Y. Yu, Z. Wang, G. Shao, *J. Mater. Chem. A* **2018**, 6, 19843.

[81] Y. Yu, Z. Wang, G. Shao, *J. Mater. Chem. A* **2019**, 7, 21985.

[82] Z. Deng, Z. Zhu, I.-H. Chu, S. P. Ong, *Chem. Mater.* **2016**, 29, 281.

[83] M. V. Agnihotri, S.-H. Chen, C. Beck, S. J. Singer, *J. Phys. Chem. B* **2014**, 118, 8170.

[84] Y. Yu, Z. Wang, G. Shao, *J. Mater. Chem. A* **2019**, 7, 10483.

[85] J. Heyd, G. E. Scuseria, M. Ernzerhof, *J. Chem. Phys.* **2003**, 118, 8207.

[86] J. Heyd, G. E. Scuseria, *J. Chem. Phys.* **2004**, 121, 1187.

[87] J. Heyd, G. E. Scuseria, M. Ernzerhof, *J. Chem. Phys.* **2006**, 124, 219906.

**Figure 1.** a) The convex hull of $Li_6PS_{5(1-x)}O_{5x}Cl$ ($0 \le x \le 1$) against S/O ratio. The configurations for labeled compounds of interest: b) $Li_6PS_{4.25}O_{0.75}Cl$, c) $Li_6PS_4OCl$, and d) $Li_6PSO_4Cl$. e) The XRD patterns of $Li_{6.25}PS_{3.5}O_{1.75}Cl_{0.75}$, $Li_{6.25}PS_4O_{1.25}Cl_{0.75}$, $Li_{6.25}PS_{4.2}O_{1.05}Cl_{0.75}$, and $Li_6PS_5Cl$. f) The enlarged XRD patterns from 29 to 32°. g) Raman spectrum of pristine $Li_6PS_5Cl$ (dark green) and $Li_{6.25}PS_4O_{1.25}Cl_{0.75}$ (orange). h) The enlarged Raman spectrum for the main peak.

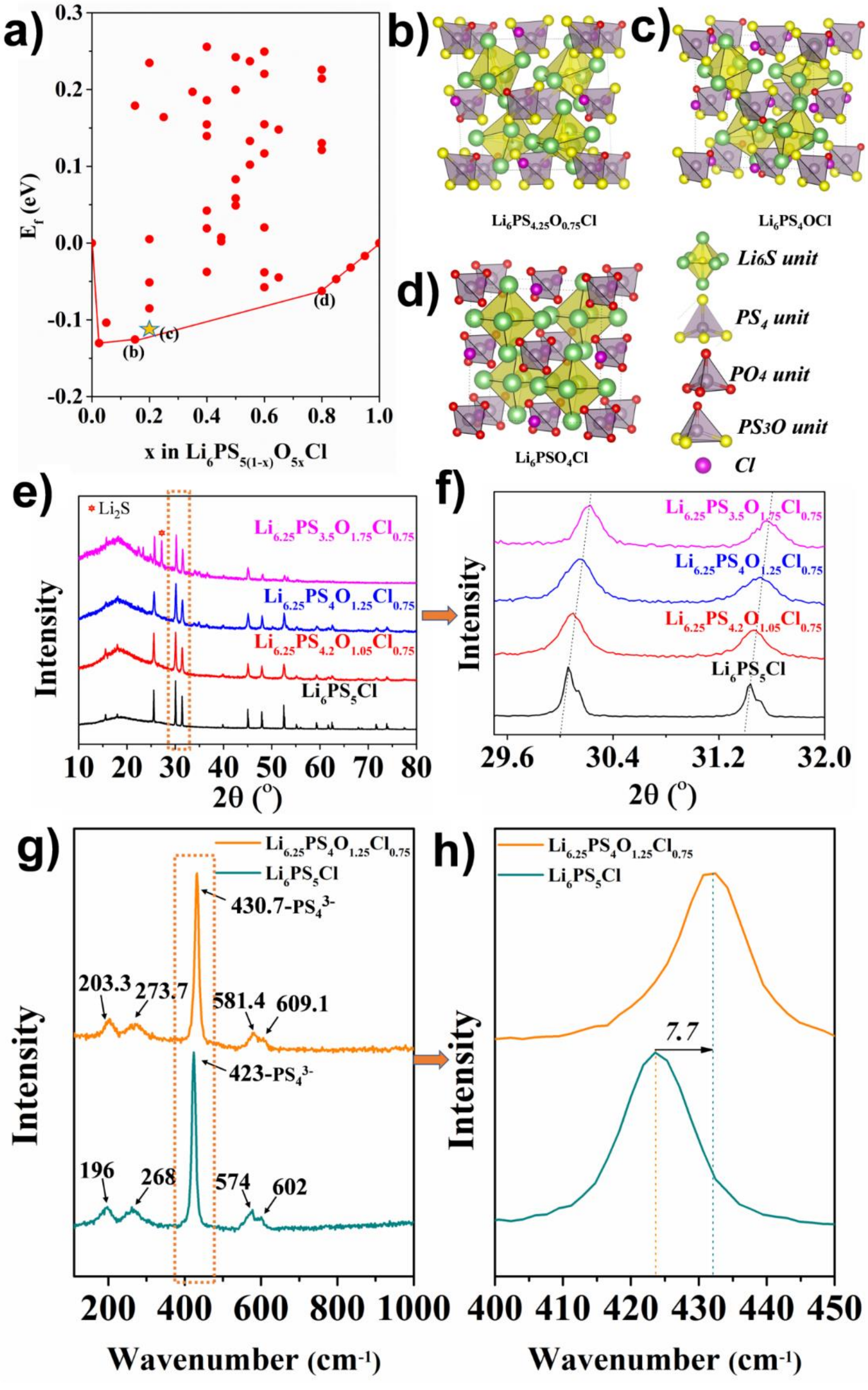

**Figure 2.** Impedance spectra over a broad frequency range from 100 Hz to 7 MHz carried out at various temperatures from 35 to 95 °C for a) $Li_6PS_5Cl$ and b) $Li_{6.25}PS_4O_{1.25}Cl_{0.75}$. Inset shows typical equivalent circuit. Impedance spectra of c) $Li_6PS_5Cl$ and d) $Li_{6.25}PS_4O_{1.25}Cl_{0.75}$ at low temperature from 0 to -20 °C. e) Ionic conductivity for $Li_6PS_5Cl$ (black) and $Li_{6.25}PS_4O_{1.25}Cl_{0.75}$ (red) from experiments (color dots) and AIMD simulation (dotted line) plotted as Log (σT) vs. (1000/T) based on the Nernst–Einstein equation. Densities of AIMD trajectories at 1000 K for f) $Li_6PS_5Cl$ and g) $Li_{6.25}PS_4O_{1.25}Cl_{0.75}$.

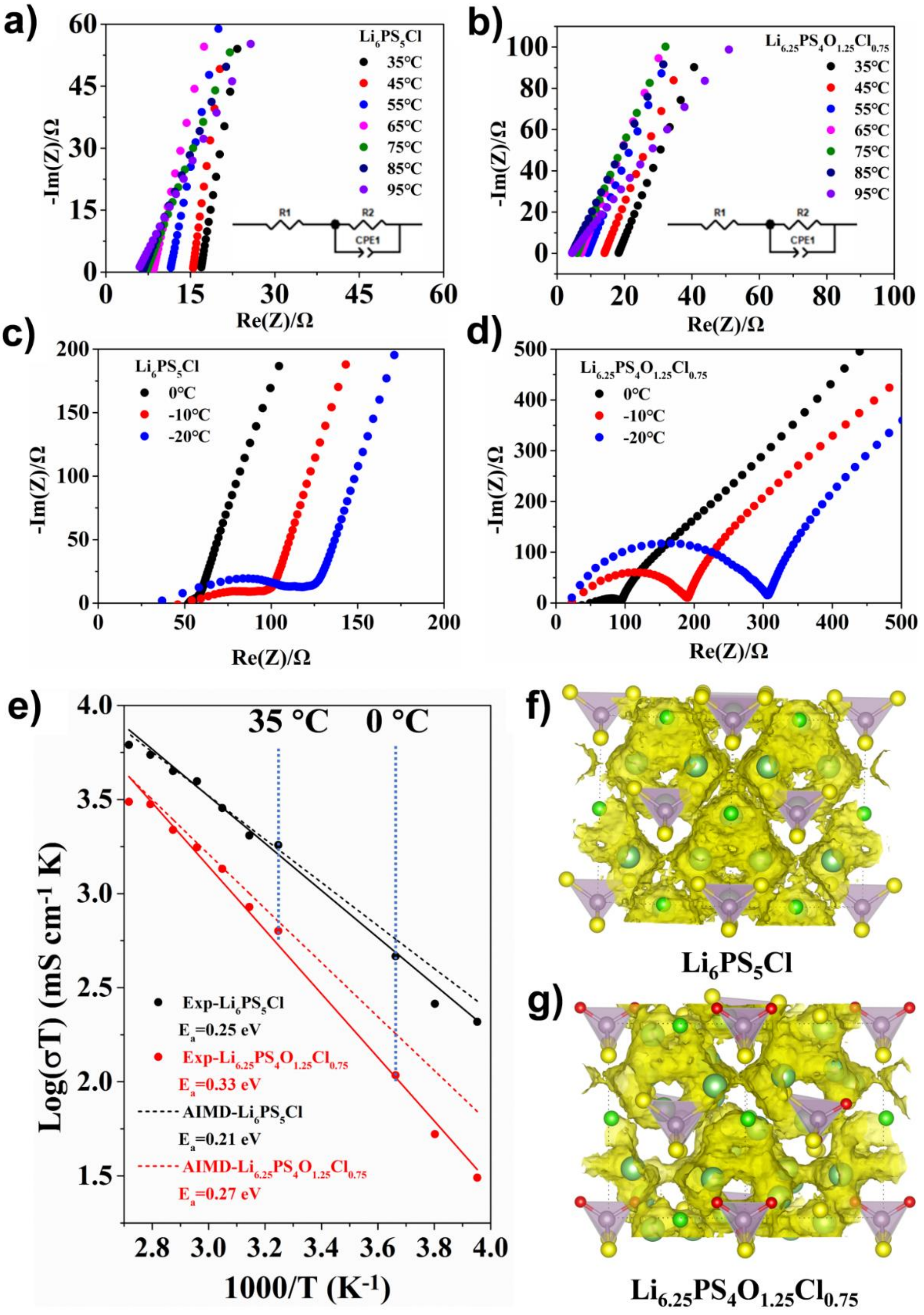

**Figure 3.** a) DSC measurements of $Li_6PS_5Cl$ (black) and $Li_{6.25}PS_4O_{1.25}Cl_{0.75}$ (red) from room temperature to 350 °C. b) XRD patterns of $Li_6PS_5Cl$ and $Li_{6.25}PS_4O_{1.25}Cl_{0.75}$ after being exposed to air with humidity of 53% for 0.5 h. c) XRD patterns of the post-annealed sample $Li_{6.25}PS_4O_{1.25}Cl_{0.75}$, $Li_6PS_5Cl$ and the pristine phases, with stars from the impurity $Li_2S$. d) Impedance spectra for the post-annealed sample $Li_{6.25}PS_4O_{1.25}Cl_{0.75}$ and $Li_6PS_5Cl$. Inset shows the enlarged impedance spectra at high frequency.

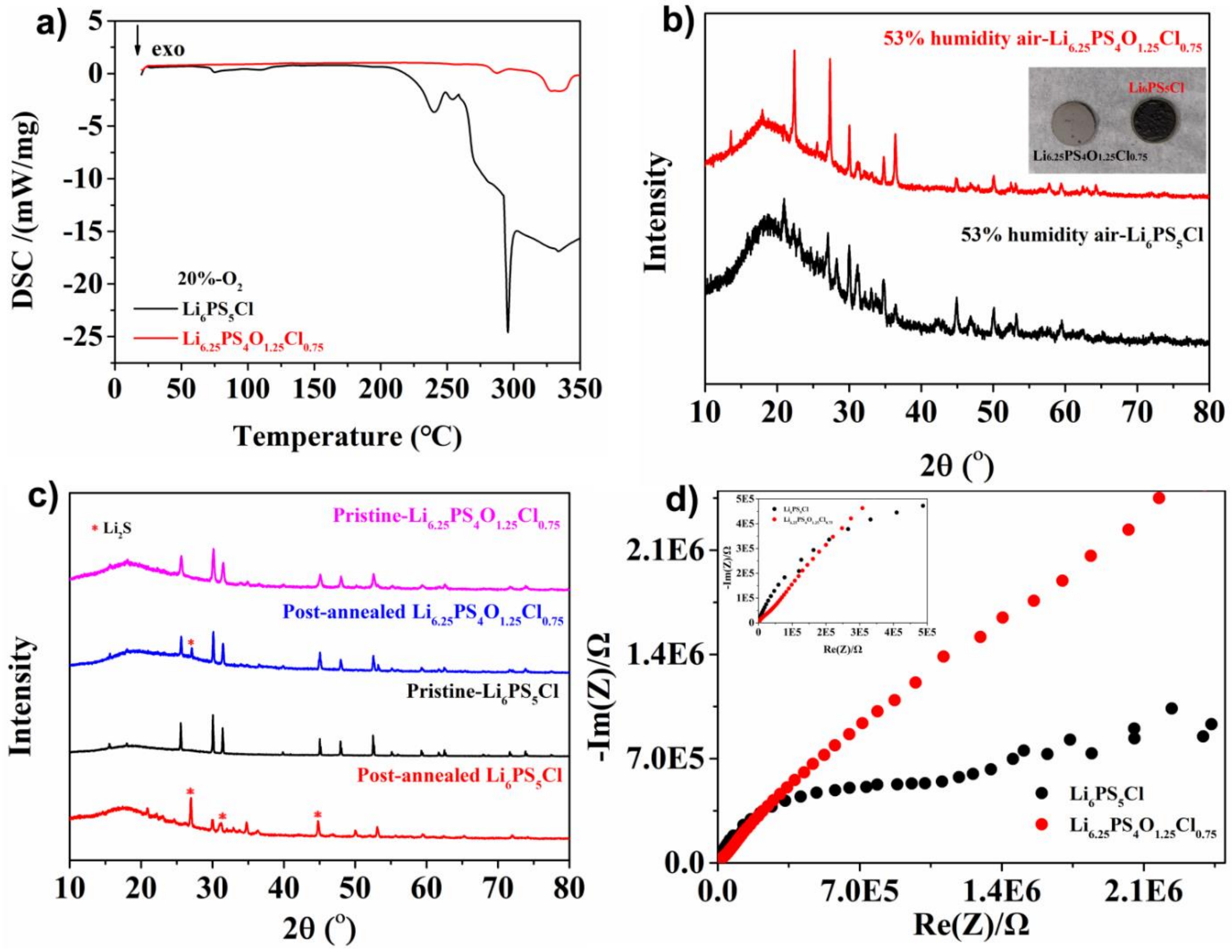

**Figure 4.** a) Galvanostatic cycling of Li|$Li_6PS_5Cl$|Li (green) and Li|$Li_{6.25}PS_4O_{1.25}Cl_{0.75}$|Li (orange) symmetric cells with various current density from 0.05 to 0.5A/ $cm^2$ for 200 h. b) Galvanostatic cycling of Li|$Li_{6.25}PS_4O_{1.25}Cl_{0.75}$|Li with various current density from 0.55 to 1A/ $cm^2$. c) The prolonged cycling of Li|$Li_{6.25}PS_4O_{1.25}Cl_{0.75}$|Li at 1 mA/$cm^2$ with a cut-off capacity of 1 mAh $cm^{-2}$ for another 1400 h. Insets indicating the magnified regions of the voltage profile at different time.

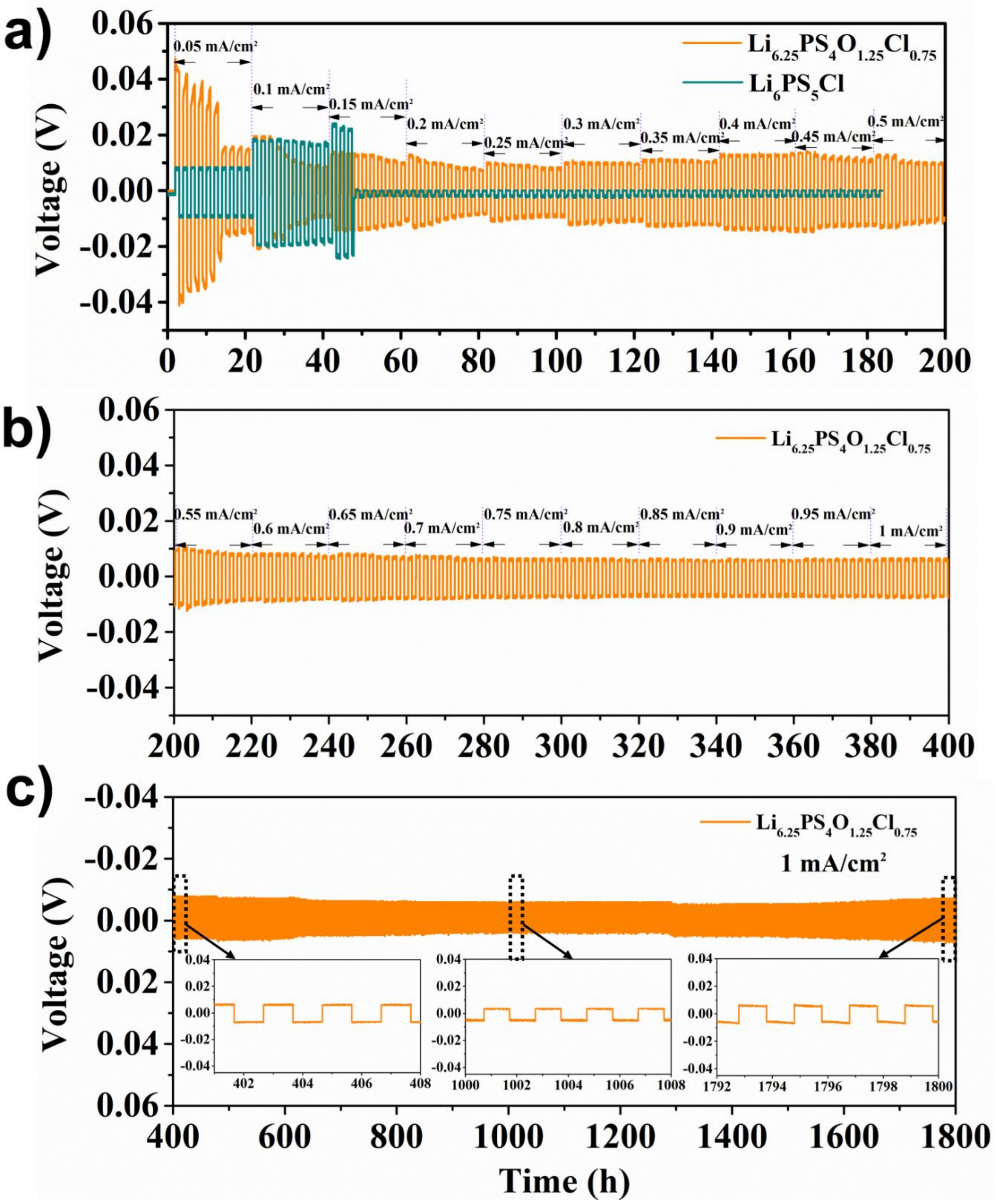

**Figure 5.** Raman spectrum for the collected samples at the interface of a) Li|$Li_6PS_5Cl$, after cycling at the current density of 0.15 mA $cm^{-2}$ for 10 h and b) Li|$Li_{6.25}PS_4O_{1.25}Cl_{0.75}$, after cycling at the current density of 1 mA/$cm^2$ for 50 h. c) The schematic diagram for the growth of Li dendrites at the interface of Li|$Li_6PS_5Cl$, and d) schematic diagrams showing the mechanism for suppression of Li dendrites at the interface of Li|$Li_{6.25}PS_4O_{1.25}Cl_{0.75}$.

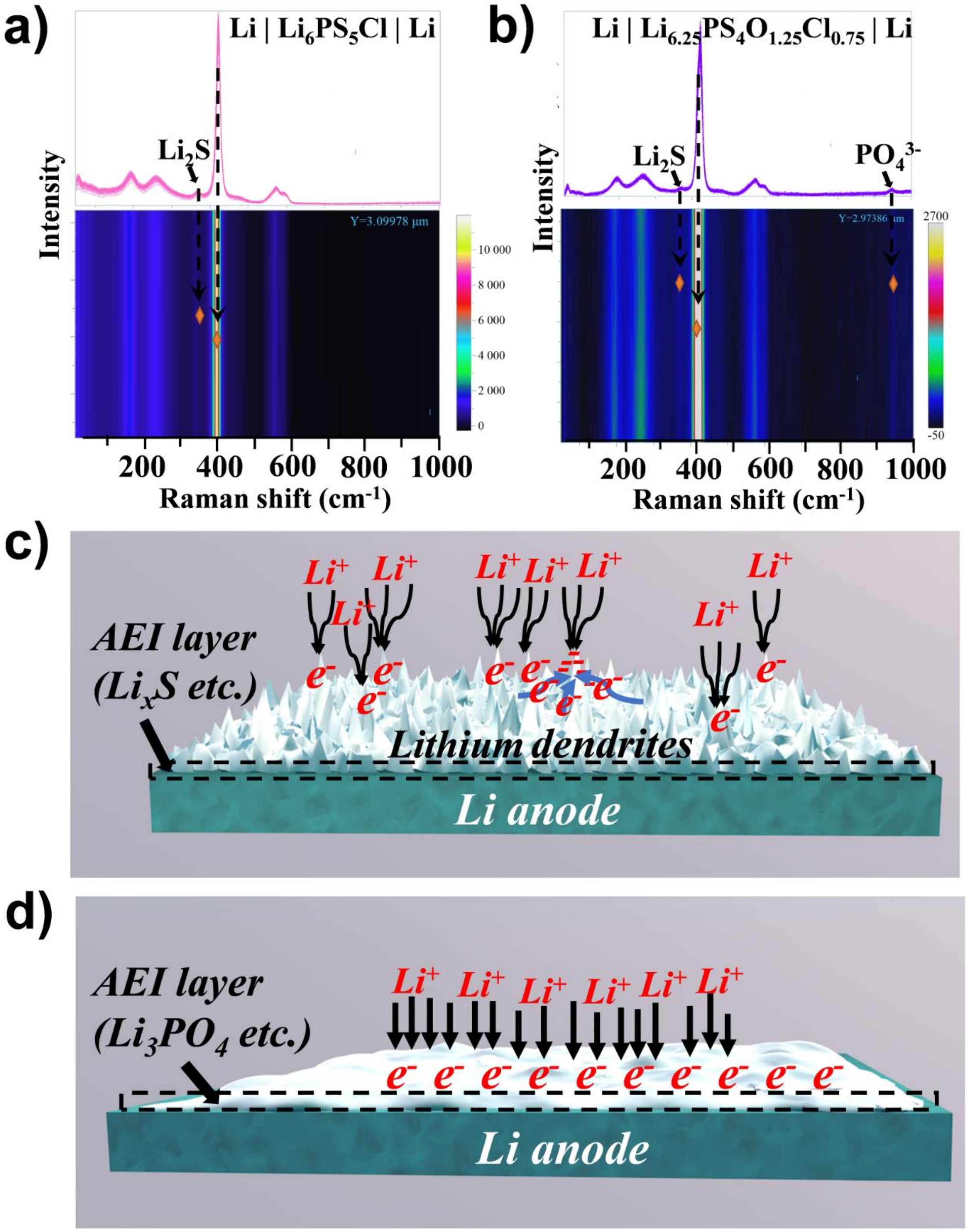

**Figure 6.** Theoretical electronic voltage windows for a) $Li_6PS_5Cl$ and b) $Li_6PS_4OCl$ based on equilibrium phase simulations. c) Experimental cyclic voltammetry (CV) tests for the cells in the form of C+$Li_6PS_5Cl$|$Li_6PS_5Cl$|Li (green), and C+$Li_{6.25}PS_4O_{1.25}Cl_{0.75}$|$Li_{6.25}PS_4O_{1.25}Cl_{0.75}$|Li (orange).

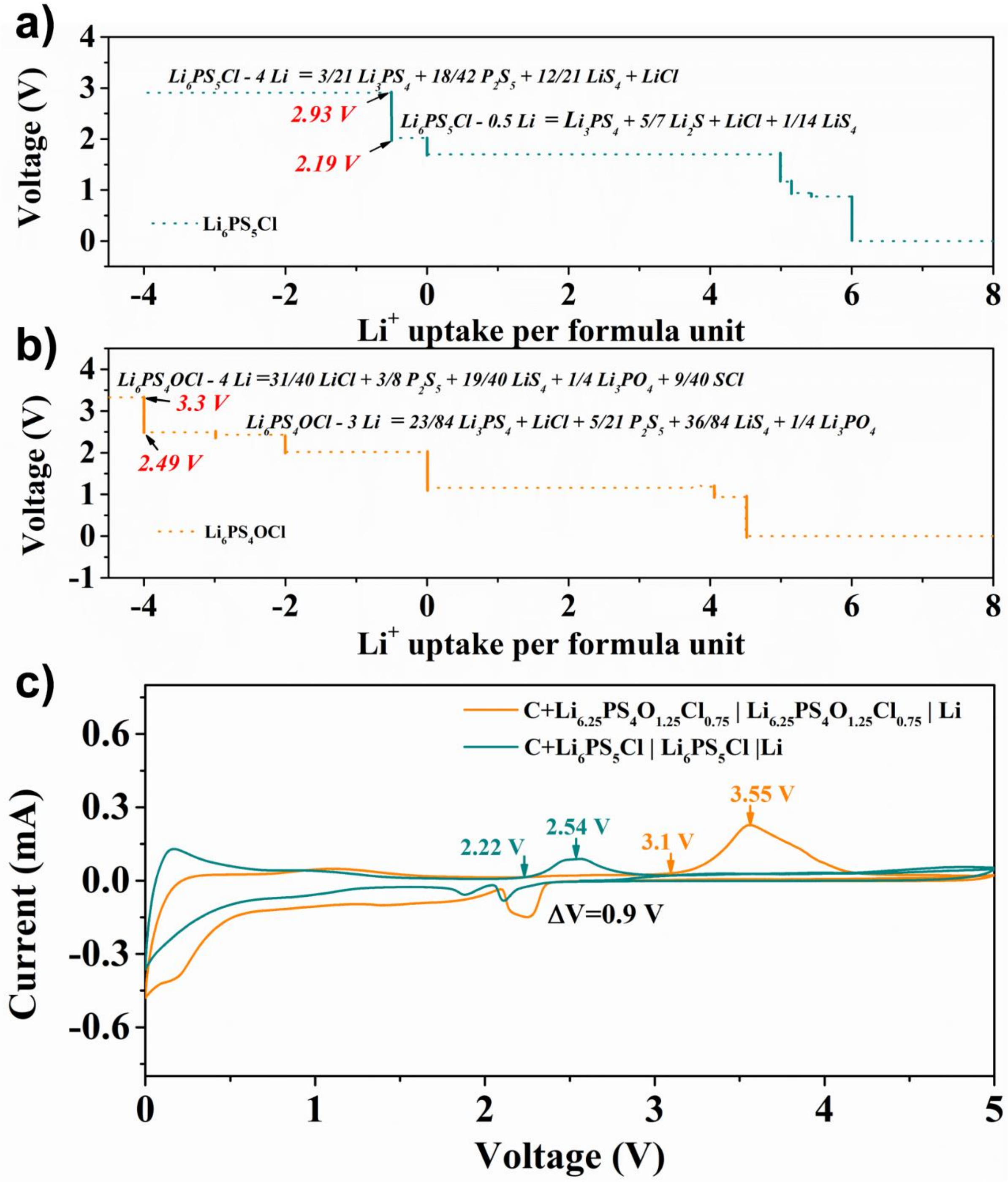

**Figure 7.** a) Schematic diagram showing the process for the fabrication of coin cells. ASSB cells of $LiCoO_2|Li_6PS_5Cl|Li$ and $LiCoO_2|Li_{6.25}PS_4O_{1.25}Cl_{0.75}|Li$ (with active $LiCoO_2$ cathode loading of 4 mg cm$^{-2}$) are assembled in the same way. b) The charge/discharge voltage profiles of ASSBs at 0.05 mA cm$^{-2}$ during the 1$^{st}$, 2$^{nd}$, and 10$^{th}$ cycling. c) The cycling performance of ASSBs. Raman tests from the generated CEI at the interface of d) $LiCoO_2|Li_6PS_5Cl$ and e) $LiCoO_2|Li_{6.25}PS_4O_{1.25}Cl_{0.75}$ in the ASSBs (Charge to 4.2 V after 10 cycles).

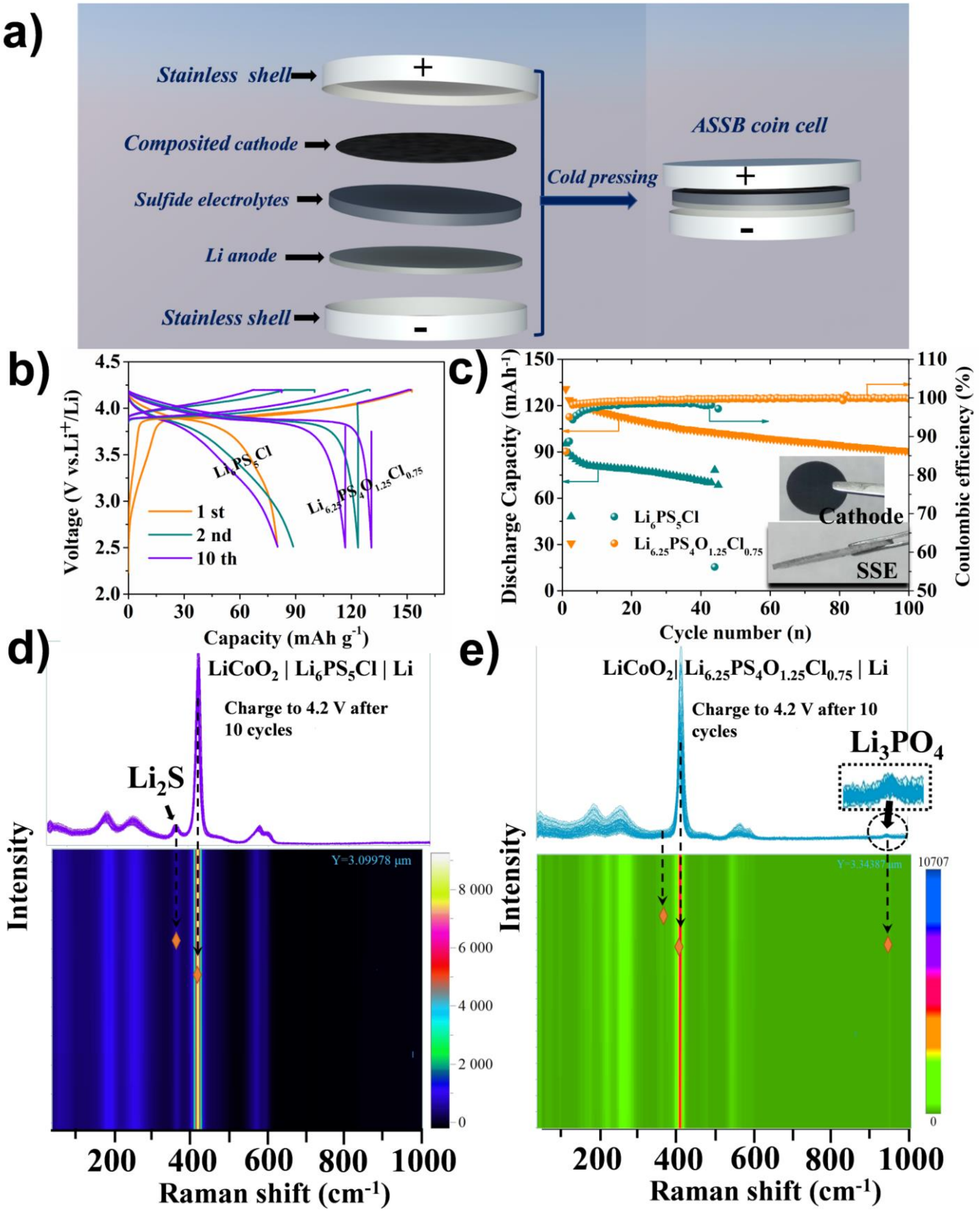

**Figure 8.** XPS spectra of the composite cathode of $LiCoO_2|Li_6PS_5Cl|Li$: a) sulfur 2p region and b) phosphorus 2p region before cycling; c) sulfur 2p region and d) phosphorus 2p region after 10 cycles charged to 4.2 V. XPS spectra of the composited electrode of $LiCoO_2|Li_{6.25}PS_4O_{1.25}Cl_{0.75}|Li$, e) sulfur 2p region and f) phosphorus 2p region before cycling, g) sulfur 2p region and h) phosphorus 2p region before cycling. i) Phosphorus 2p region in $Li_3PO_4$ as a standard reference. Schematic diagram for j) the depletion layer generated at the interface of $LiCoO_2|Li_6PS_5Cl$, k) the in situ generated $Li_3PO_4$ as a buffer layer at the interface of $LiCoO_2|Li_{6.25}PS_4O_{1.25}Cl_{0.75}$.

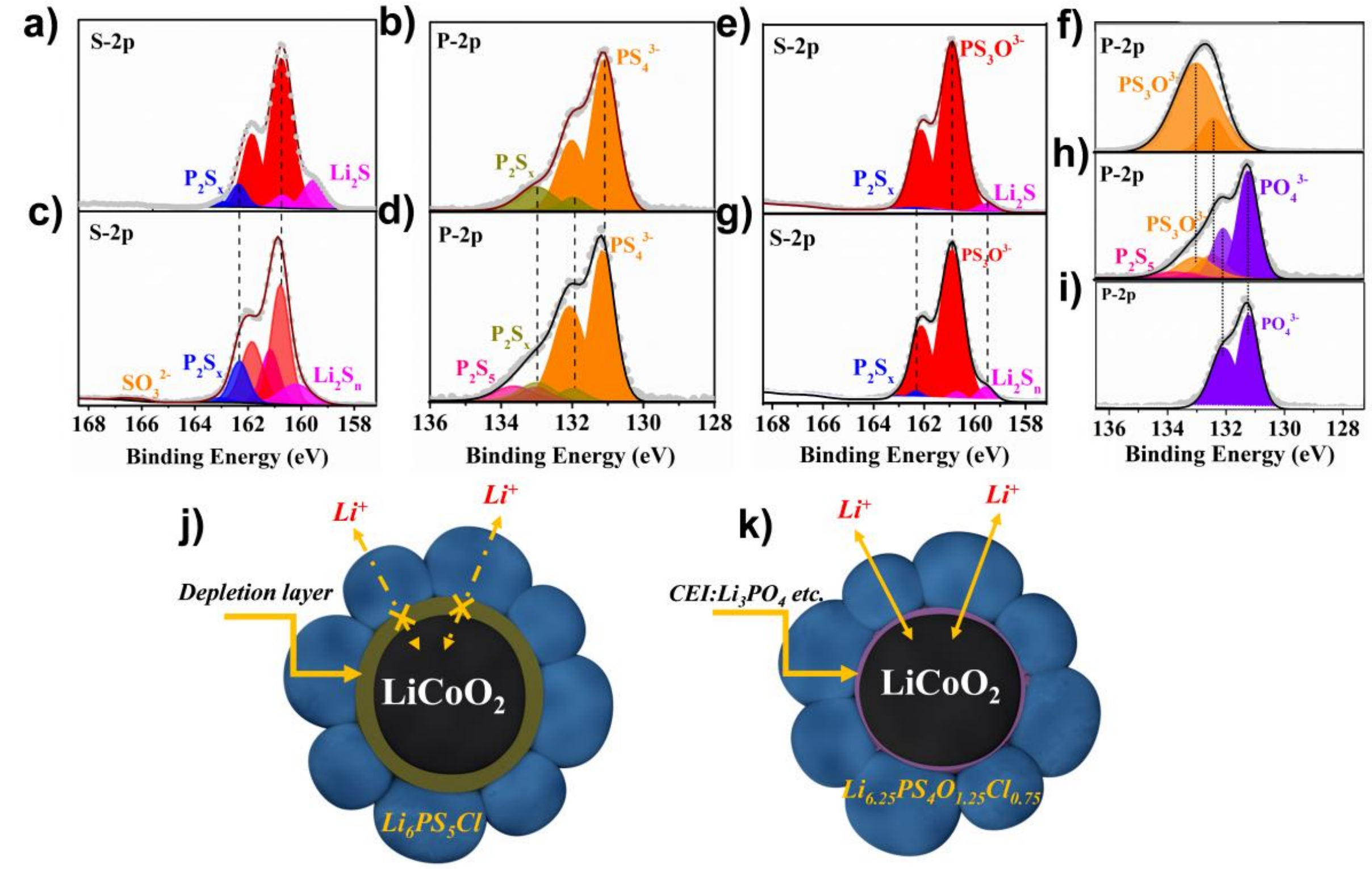

**An Effective Strategy to Enable Argyrodite Sulfides as Superb Solid-State Electrolytes: Safeguarding Remarkable Ionic Conductivity and Interfacial Stability with Electrodes**

ToC figure

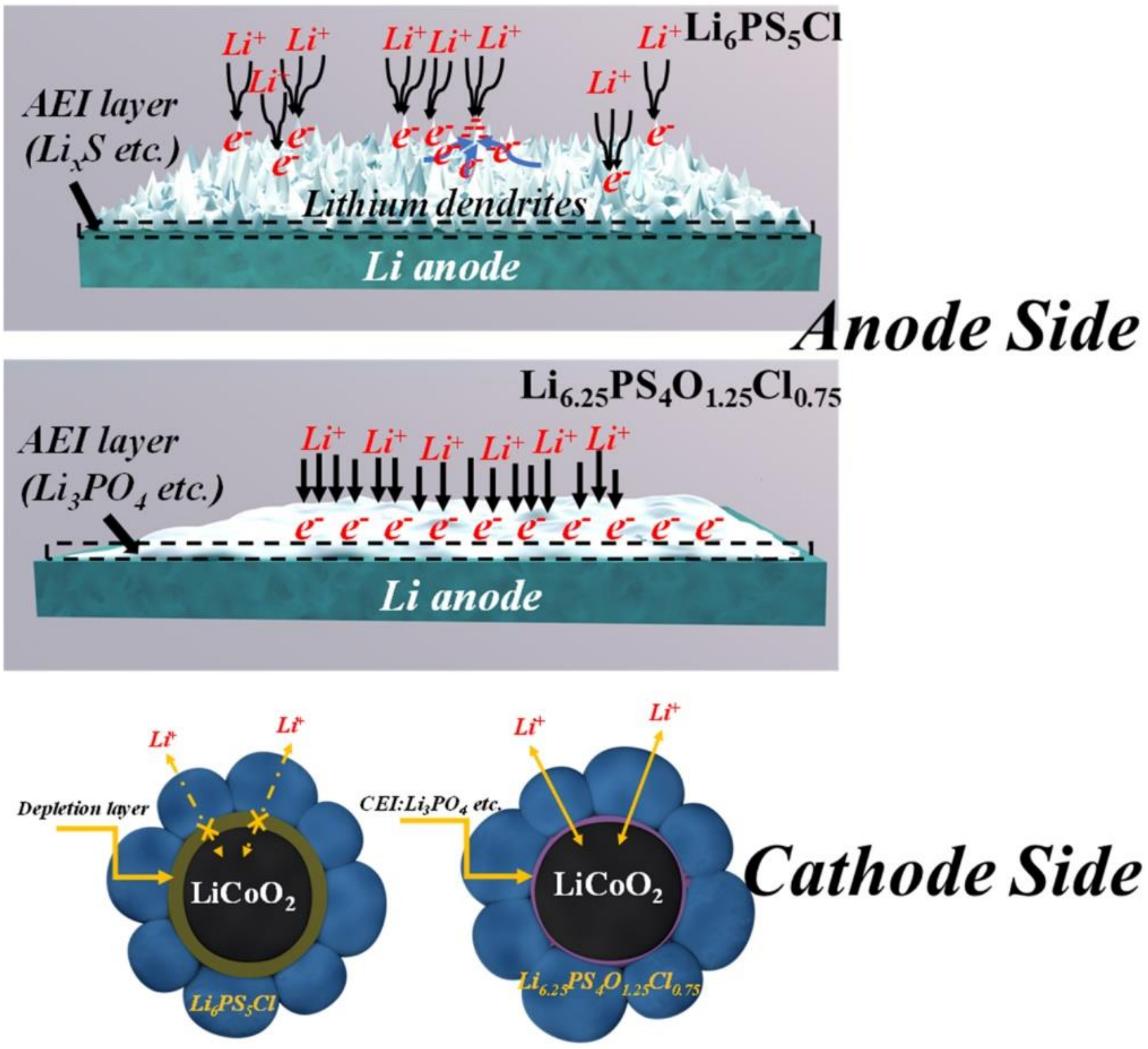

# Supporting Information (SI-)

## An Effective Strategy to Enable Argyrodite Sulfides as Superb Solid-State Electrolytes: Safeguarding Remarkable Ionic Conductivity and Interfacial Stability with Electrodes

*Hongjie Xu[1,2,3], Yuran Yu[1,2,3], Junhua Hu[1,2,3]*, Zhuo Wang[1,2,3]*, and Guosheng Shao[1,2,3]**

1 School of Materials Science and Engineering, Zhengzhou University, Zhengzhou 450001, Henan, China

2 State Centre for International Cooperation on Designer Low-carbon &Environmental Materials (CDLCEM), Zhengzhou University, Zhengzhou 450001, Henan, China

3 Zhengzhou Materials Genome Institute (ZMGI), Zhengzhou 450100, Henan, China

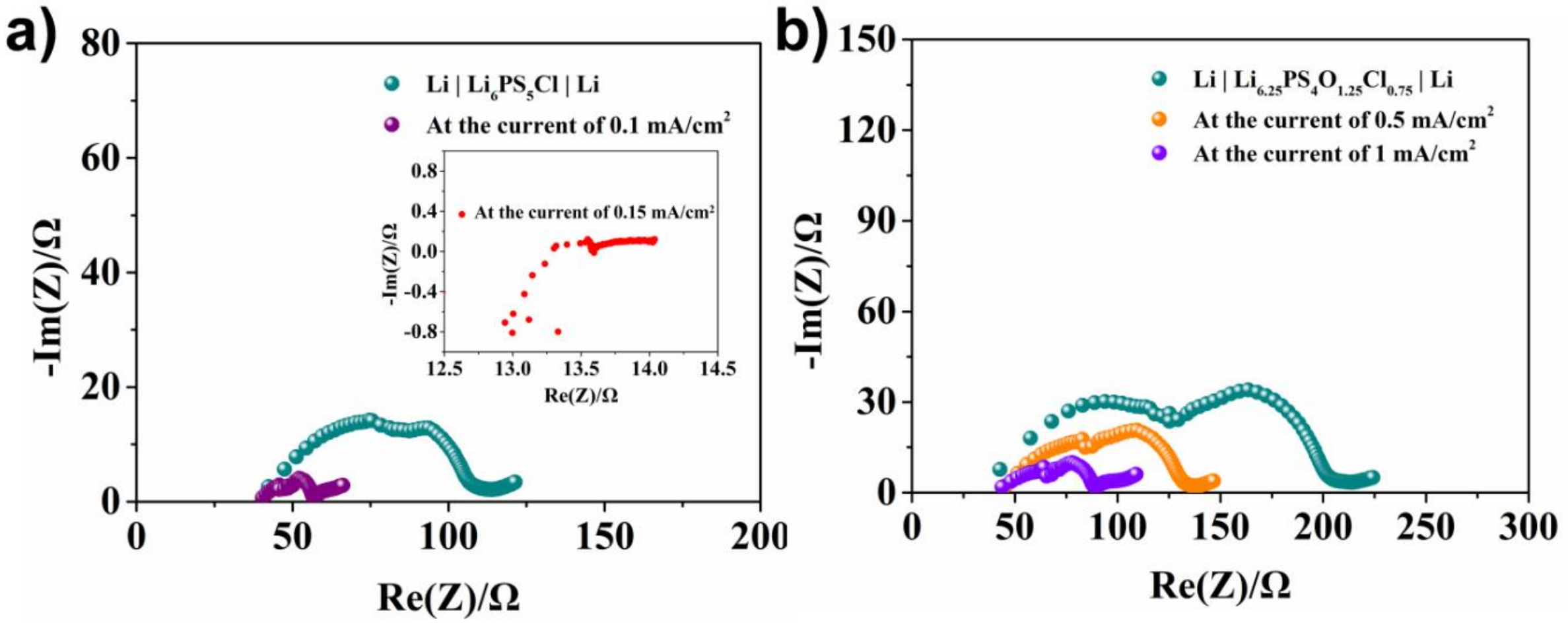

Figure S1: Impedance spectra for a) Li|$Li_6PS_5Cl$|Li and b) Li|$Li_{6.25}PS_4O_{1.25}Cl_{0.75}$|Li symmetric cells before cycling and after cycling at different current densities. And, inset in a) shows the noisy spectrum at 0.15 mA/$cm^2$.

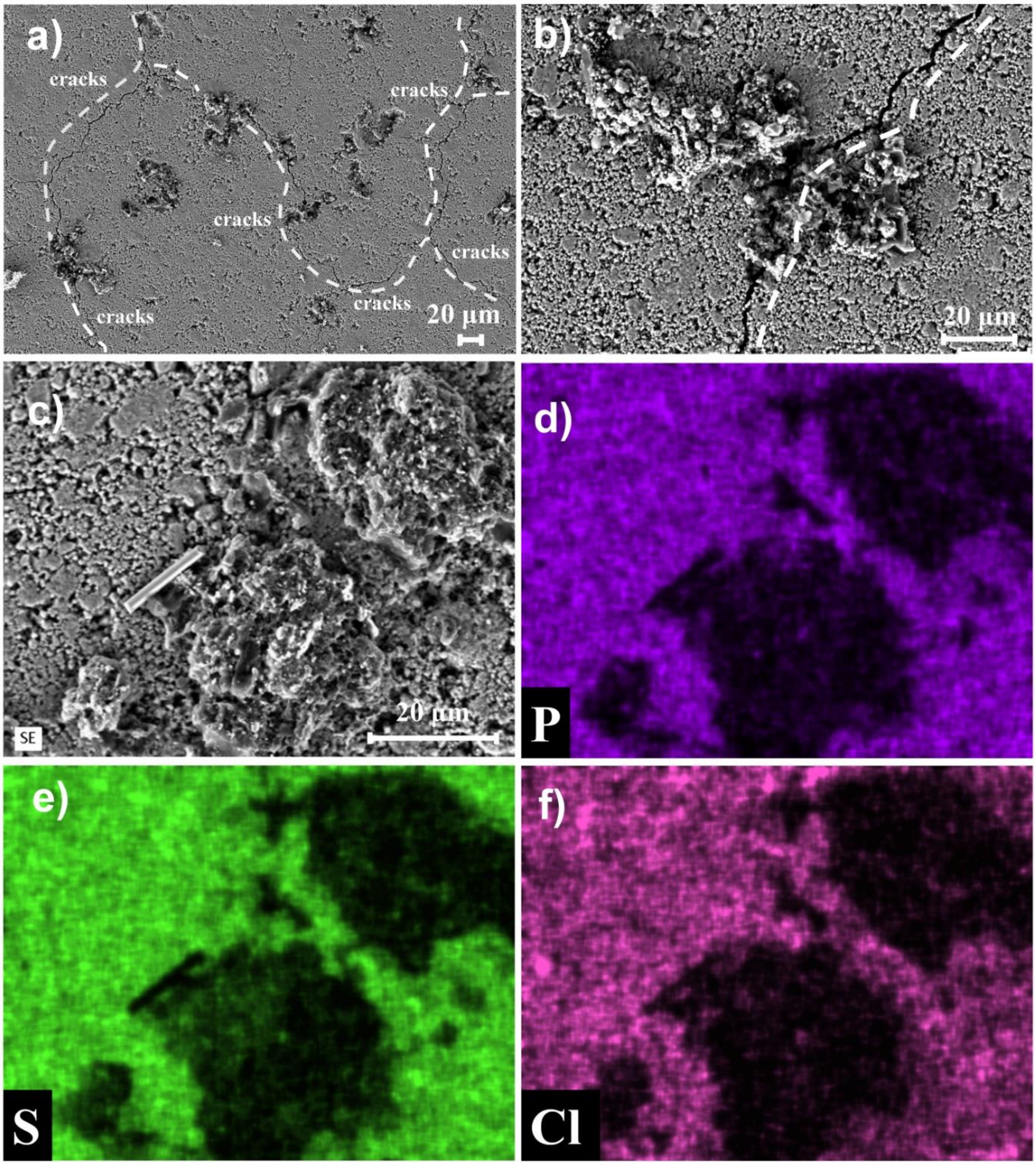

Figure S2: SEM image for the top view of $Li_6PS_5Cl$ pellet disassembled from the symmetric cell Li|$Li_6PS_5Cl$|Li after plating and stripping with a current density of 0.15 mA $cm^{-2}$ for 10 h. a) The overview of lithium dendrites with a lot of cracks stretch inside of pellet (white solid line). b) The enlarged view of lithium dendrites and cracks existing on the pellet. c) SEM image corresponding to the EDS mapping of d) P, e) S, and f) Cl element. The black areas correspond to the Li element.

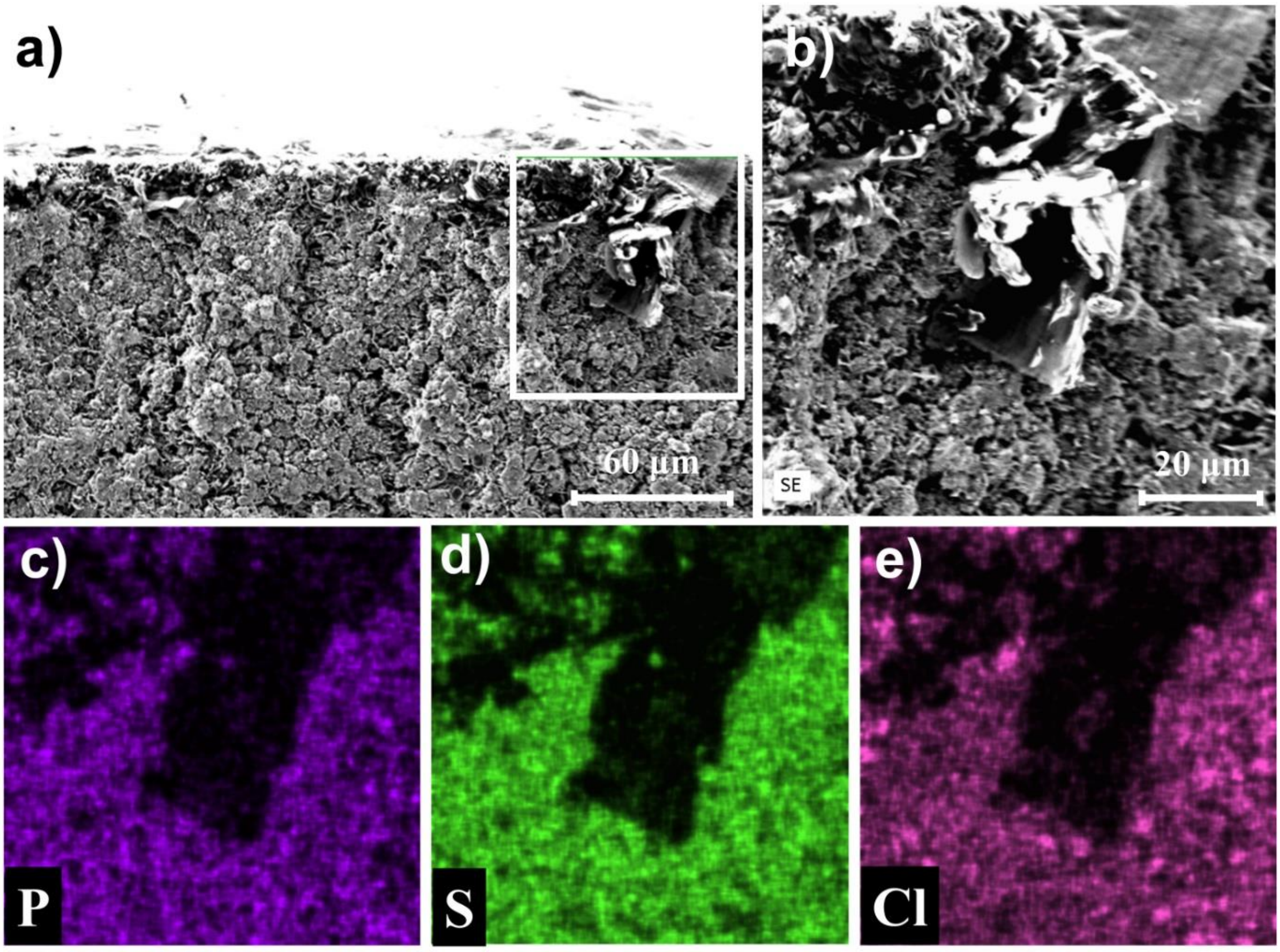

Figure S3: a) SEM image for the cross-section view of $Li_6PS_5Cl$ pellet disassembled from the symmetric cell Li|$Li_6PS_5Cl$|Li after plating and stripping at a current density of 0.15 mA $cm^{-2}$ for 10 h. b) The enlarged view of lithium dendrites grown across the cross-section of pellet. c)-e) EDS mapping of P, S, and Cl element, correspondingly. The black areas indicate the distribution of Li element.

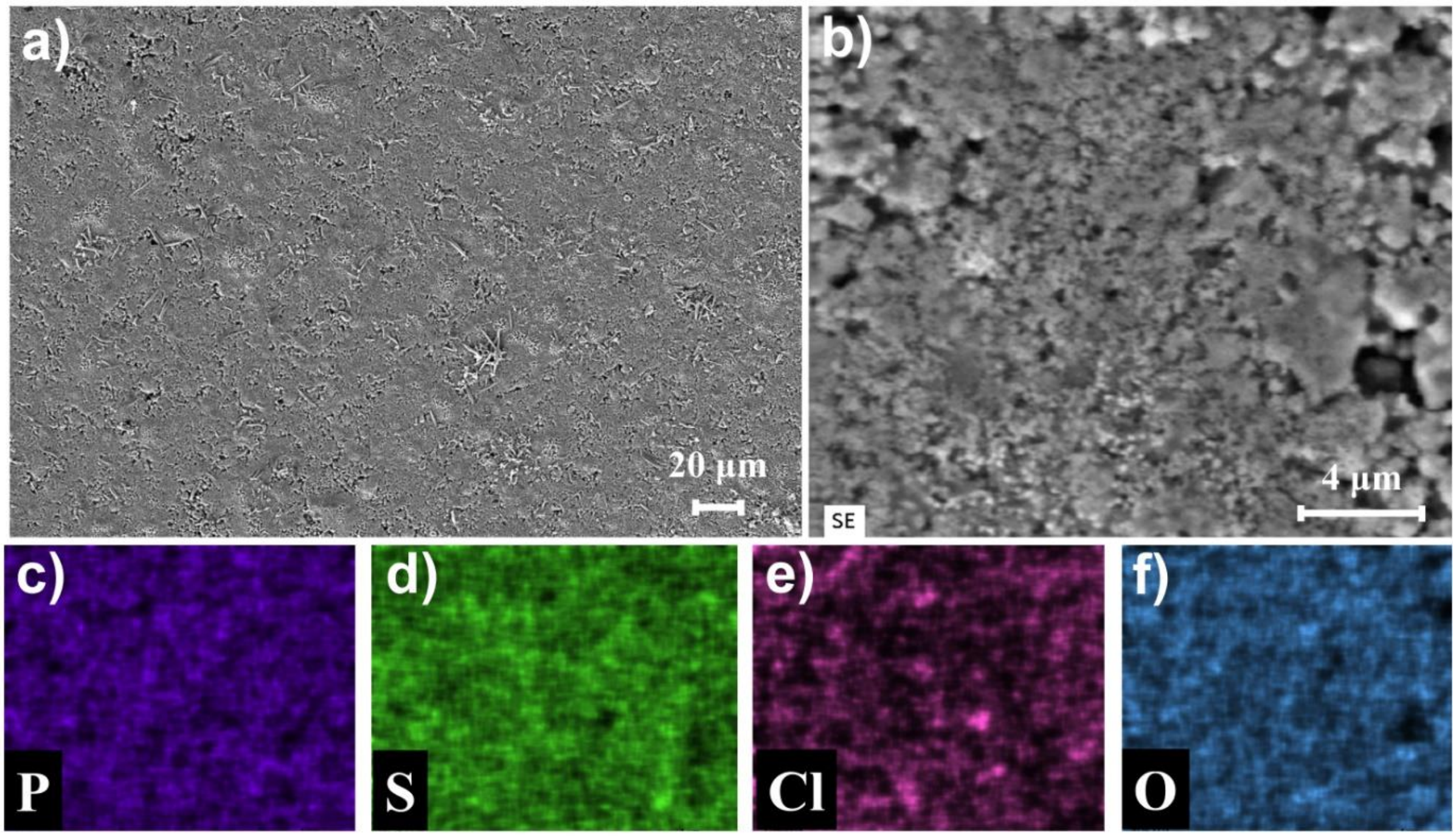

Figure S4: a) SEM image for the cross-section view of $Li_6PS_5Cl$ pellet disassembled from SEM image for the top view of $Li_{6.25}PS_4O_{1.25}Cl_{0.75}$ pellet disassembled from the symmetric cell Li| $Li_{6.25}PS_4O_{1.25}Cl_{0.75}$|Li after plating and stripping at a current density of 1 mA $cm^{-2}$ for 50 h. b) EDS image, and EDS mapping of c) P, d) S, e) Cl and f) O element.

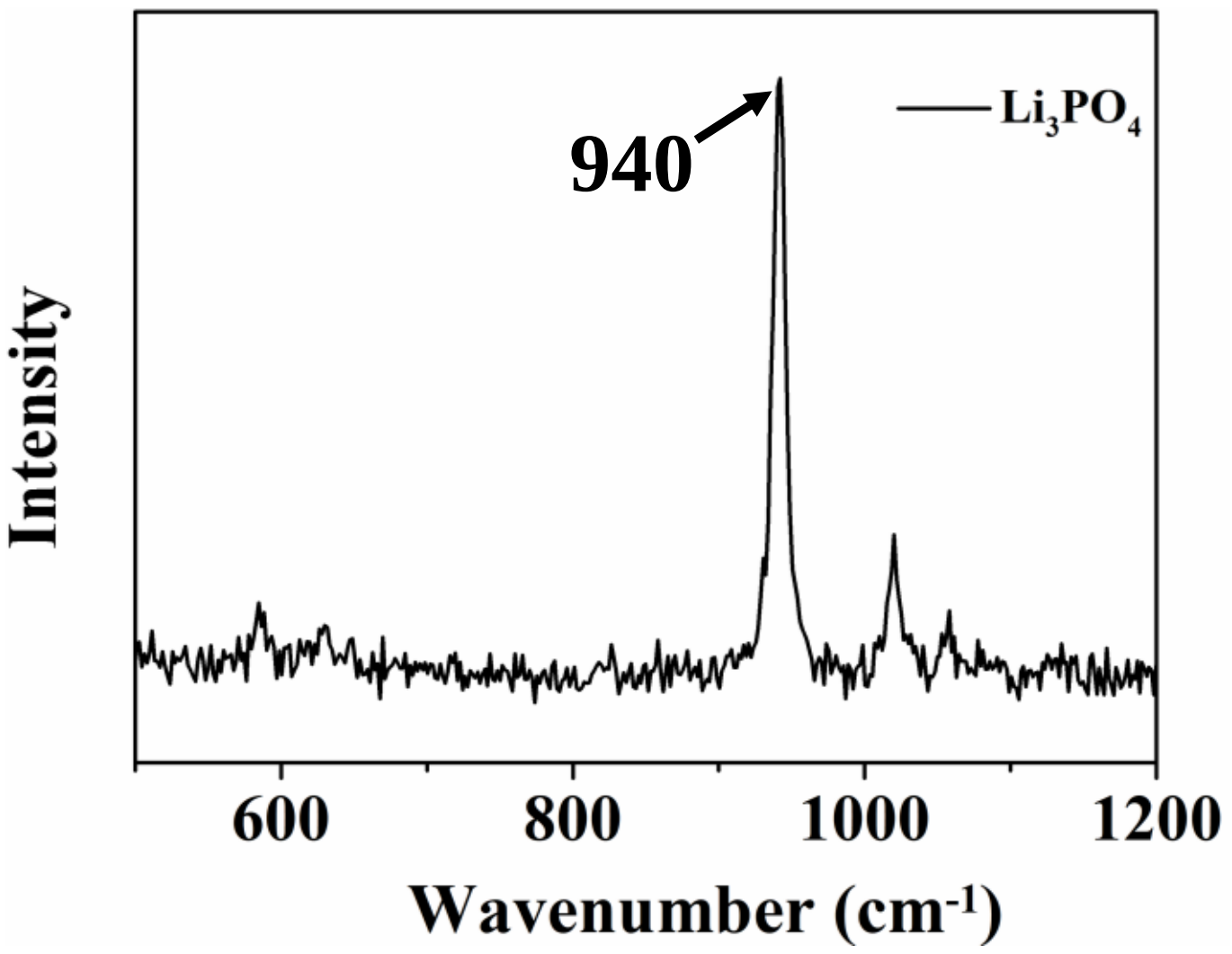

Figure S5: Raman spectrum for the reference sample of $Li_3PO_4$, the main peak at 940 $cm^{-1}$ corresponding to vibrational modes of $PO_4^{3-}$.

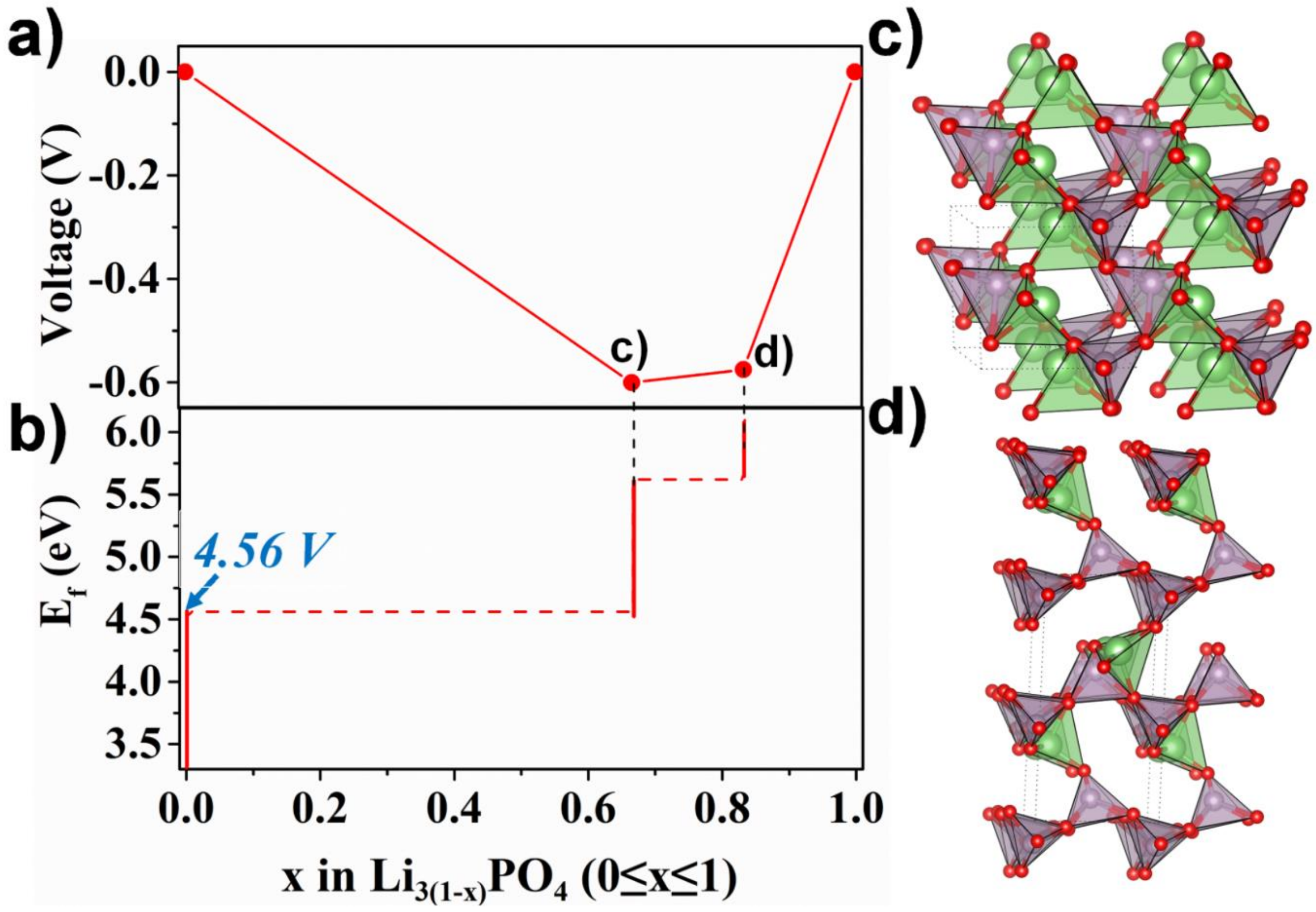

Figure S6: a) The convex hull for $Li_{3(1-x)}PO_4$ ($0 \leq x \leq 1$) through ATAT simulation with reference to the stable states of $Li_3PO_4$ and $PO_4$. b) Voltage plateaus. The configurations for some stable states of $Li_{3(1-x)}PO_4$ ($0 \leq x \leq 1$), c) $LiPO_4$, d) $Li_{0.5}PO_4$.

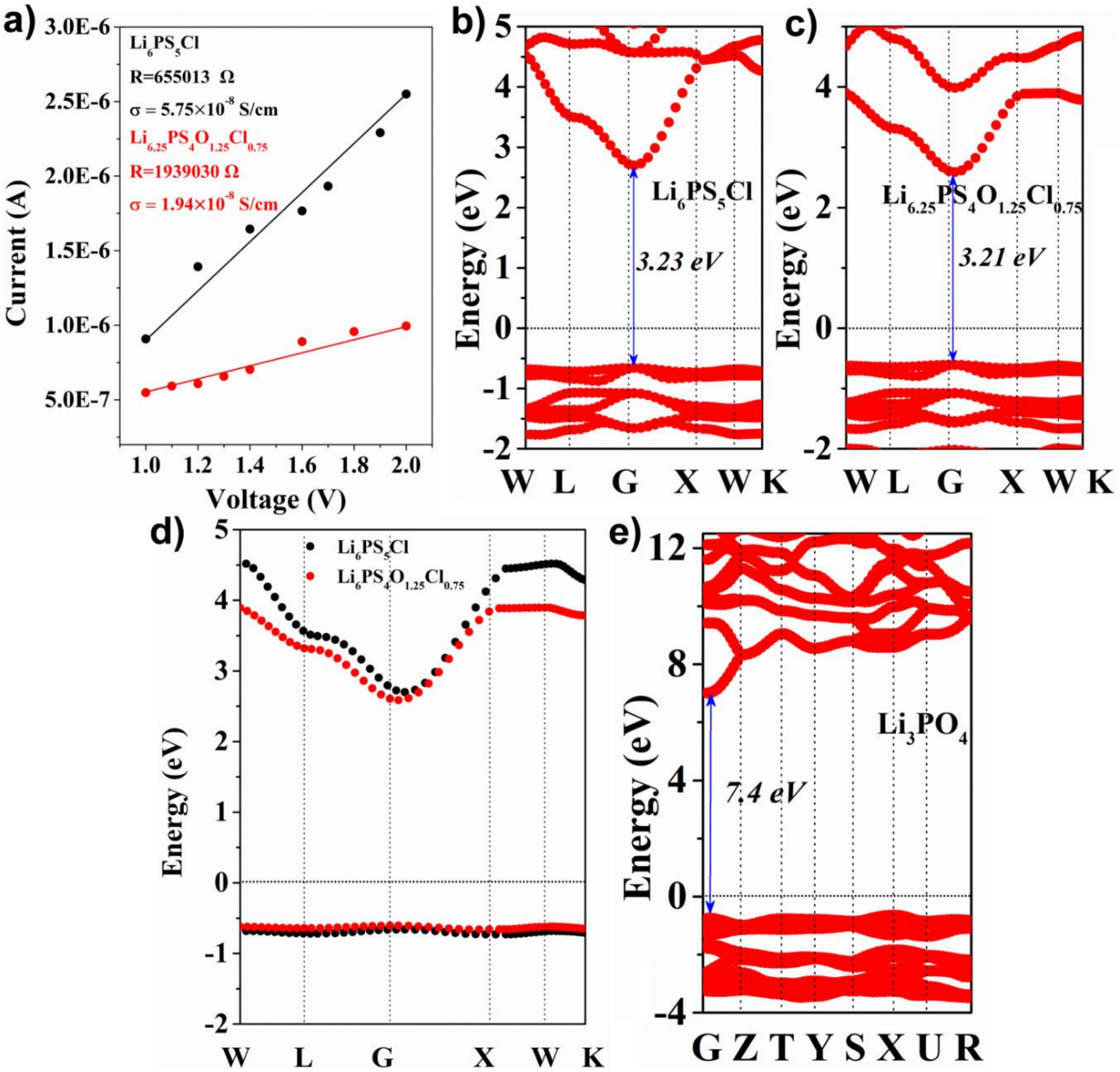

Figure S7: a) The electronic conductivities of $Li_6PS_5Cl$ and $Li_{6.25}PS_4O_{1.25}Cl_{0.75}$ by the direct current (DC) polarization method, The electronic band structure of b) $Li_6PS_5Cl$, c) $Li_{6.25}PS_4O_{1.25}Cl_{0.75}$ with d) a direct comparison for their conduction band minimum (CBM), and e) $Li_3PO_4$.

**Table S1.** Lattice parameters and symmetry information of $Li_6PS_5Cl$, $Li_{6.25}PS_4O_{1.25}Cl_{0.75}$, and $Li_{6.25}PS_4O_{1.25}Cl_{0.75}$.

| System | a (Å) | b (Å) | c (Å) | α (°) | β (°) | γ(°) | Sym | Num |
|---|---|---|---|---|---|---|---|---|
| **$Li_6PS_5Cl$** | 9.859 | 9.859 | 9.859 | 90.0 | 90.0 | 90.0 | $F\bar{4}3m$ | 216 |
| **$Li_{6.25}PS_{4.2}O_{1.05}Cl_{0.75}$** | 9.851 | 9.851 | 9.851 | 90.0 | 90.0 | 90.0 | $F\bar{4}3m$ | 216 |
| **$Li_{6.25}PS_4O_{1.25}Cl_{0.75}$** | 9.84 | 9.84 | 9.84 | 90.0 | 90.0 | 90.0 | $F\bar{4}3m$ | 216 |

**Table S2.** The states of $Li_6PS_5Cl$ during de-lithiation or lithiation process with reference to the equilibrium phase (E.P.), the corresponding formation energies ($E_f$) and voltage plateaus are listed.

| De-lithiation/lithiation | E.P. | $E_f$ (eV/atom) | Voltage/V |
|---|---|---|---|
| $Li_6PS_5Cl$ + 8 Li | $Li_3P+5Li_2S+LiCl$ | 0.5929 | 0.0-0.872 |
| $Li_6PS_5Cl$ + 6 Li | $LiP+5Li_2S+LiCl$ | 0.5627 | 0.0-0.872 |
| $Li_6PS_5Cl$ + 5.429 Li | $0.143Li_3P_7+5Li_2S+LiCl$ | 0.5507 | 0.872-0.94 |
| $Li_6PS_5Cl$ + 5.143 Li | $0.143LiP_7+5Li_2S+LiCl$ | 0.5409 | 0.94-1.166 |
| $Li_6PS_5Cl$ + 5 Li | $P+5Li_2S+LiCl$ | 0.5354 | 1.166-1.7 |
| $Li_6PS_5Cl$ + 0 Li | $Li_3PS_4+Li_2S+LiCl$ | 0.0761 | 1.7-2.02 |
| $Li_6PS_5Cl$ – 0.5Li | $Li_3PS_4$+ $5/7Li_2S+LiCl+1/14LiS_4$ | 0.0744 | 2.02-2.19 |
| $Li_6PS_5Cl$ – 4Li | $3/21Li_3PS_4+18/42P_2S_5+12/21LiS_4+LiCl$ | 0.042 | 2.19-2.93 |
| $Li_6PS_5Cl$ – 5.5Li | $SCl+0.5P_2S_5+0.375LiS_4$ | 0.084 | 2.93 |
| $Li_6PS_5Cl$ – 6Li | $1/2P_2S_5+3/2S+SCl$ | 1.007 | 14.42493 |

**Table S3.** The states of $Li_6PS_4OCl$ during de-lithiation or lithiation process with reference to the equilibrium phase (E.P.), the corresponding formation energies ($E_f$) and voltage plateaus are also listed.

| **Delithiation/lithiation** | **E.P.** | **$E_f$ (eV/atom)** | **Voltage/V** |
|---|---|---|---|
| $Li_6PS_4OCl$ + 8 Li | $Li_3P$+4$Li_2S$+LiCl+$Li_2O$ | 0.165 | 0-0.938 |
| $Li_6PS_4OCl$ + 4.5 Li | 0.75LiP+4$Li_2S$+LiCl+0.25$Li_3PO_4$ | 0.4629 | 0-0.938 |
| $Li_6PS_4OCl$ + 4.07 Li | 0.107$Li_3P_7$+4$Li_2S$+LiCl+0.25$Li_3PO_4$ | 0.45073 | 0.938-1.183 |
| $Li_6PS_4OCl$ + 3.857 Li | 0.107$LiP_7$+4$Li_2S$+LiCl+0.25$Li_3PO_4$ | 0.44137 | 1.183-1.21 |
| $Li_6PS_4OCl$ + 3.75 Li | 0.75P+4$Li_2S$+LiCl+0.25$Li_3PO_4$ | 0.436423 | 1.21-1.16 |
| $Li_6PS_4OCl$ +0 Li | 3/4$Li_3PS_4$+LiCl+$Li_2S$+1/4$Li_3PO_4$ | 0.0634 | 1.16-2.022 |
| $Li_6PS_4OCl$ - 2Li | 55/84$Li_3PS_4$+ LiCl +1/21$P_2S_5$+ 24/84$LiS_4$+1/4$Li_3PO_4$ | 0.05 | 2.022-2.43 |
| $Li_6PS_4OCl$ - 3Li | 23/84$Li_3PS_4$+ LiCl +5/21$P_2S_5$ +36/84$LiS_4$+1/4$Li_3PO_4$ | 0.0478 | 2.43-2.49 |
| $Li_6PS_4OCl$ - 4Li | 31/40LiCl +3/8$P_2S_5$ +19/40$LiS_4$+1/4$Li_3PO_4$+9/40SCl | 0.0433 | 2.49-3.33 |
| $Li_6PS_4OCl$–5Li | 3/8$P_2S_5$+1/4$LiS_4$+1/4$Li_3PO_4$+SCl+1/8S | 0.1136 | 3.33 |